# A HIERARCHICAL INDEPENDENT COMPONENT ANALYSIS MODEL FOR LONGITUDINAL NEUROIMAGING STUDIES

By Yikai Wang and Ying Guo
for the Alzheimer's Disease Neuroimaging Initiative

*Emory University*
[‡]

In recent years, longitudinal neuroimaging study has become increasingly popular in neuroscience research to investigate disease-related changes in brain functions, to study neurodevelopment or to evaluate treatment effects on neural processing. One of the important goals in longitudinal imaging analysis is to study changes in brain functional networks across time and how the changes are modulated by subjects' clinical or demographic variables. In current neuroscience literature, one of the most commonly used tools to extract and characterize brain functional networks is independent component analysis (ICA), which separates multivariate signals into linear mixture of independent components. However, existing ICA methods are only applicable to cross-sectional studies and not suited for modelling repeatedly measured imaging data. In this paper, we propose a novel longitudinal independent component model (L-ICA) which provides a formal modeling framework for extending ICA to longitudinal studies. By incorporating subject-specific random effects and visit-specific covariate effects, L-ICA is able to provide more accurate estimates of changes in brain functional networks on both the population- and individual-level, borrow information across repeated scans within the same subject to increase statistical power in detecting covariate effects on the networks, and allow for model-based prediction for brain networks changes caused by disease progression, treatment or neurodevelopment. We develop a fully traceable exact EM algorithm to obtain maximum likelihood estimates of L-ICA. We further develop a subspace-based approximate EM algorithm which greatly reduce the computation time while still retaining high accuracy. Moreover, we present a statistical testing procedure for examining covariate effects on brain network changes. Simulation results demonstrate the advantages of our proposed methods. We apply L-ICA to ADNI2 study to investigate changes in brain functional networks in Alzheimer disease. Results from the L-ICA provide biologically insightful findings which are not revealed using existing methods.

[‡]Data used in preparation of this article were obtained from the Alzheimer's Disease Neuroimaging Initiative (ADNI) database (adni.loni.usc.edu). As such, the investigators within the ADNI contributed to the design and implementation of ADNI and/or provided data but did not participate in analysis or writing of this report. A complete listing of ADNI investigators can be found at: http://adni.loni.usc.edu/wp-content/uploads/how-to-apply/ADNI-Acknowledgement-List.pdf

**1. Introduction** Brain functional network analysis has been widely used in neuroimaging studies to reveal organization architectures of human brain. In functional imaging studies, neural activity is often captured by a series of 3-D fMRI brain images where the observed data represent the combinations of signals generated from various brain functional networks. One of the major objectives of fMRI-based network analysis is to decompose the observed series of brain images to identify underlying networks and characterize their spatial patterns and temporal dynamics. Independent component analysis (ICA) is one of the most commonly used tools for this purpose. As a special case of blind source separation, ICA decomposes observed fMRI signals into linear combinations of latent spatial source signals that are statistically as independent as possible. These latent independent components correspond to various functional networks. The popularity of the ICA method is mainly due to the following reasons. As a multivariate approach, ICA can jointly model the relationships among multiple voxels and hence provide a tool for investigating whole brain connectivity. Unlike second-order statistical methods such as PCA, ICA takes into account higher-order statistics, and the spatial statistical independence assumption of ICA is well-supported by the sparse nature in typical fMRI activation patterns (Calhoun et al., 2001a; Beckmann and Smith, 2004). Furthermore, ICA is a fully data-driven approach that does not require *a priori* temporal or spatial models. This makes ICA an important tool for analyzing resting-state fMRI where there is no experimental paradigm (Beckmann et al., 2005).

The classical ICA model was first applied to neuroimaging studies for single subject fMRI data decomposition (Mckeown et al., 1998). Some extensions referred as group ICA (Calhoun et al., 2001a) have been proposed to decompose the multiple-subject fMRI data. One commonly used group ICA framework is the temporal concatenation group ICA (TC-GICA) which stacks subjects' fMRI data in the temporal domain and then decompose the concatenated group data via ICA (Beckmann and Smith, 2005; Calhoun et al., 2001a; Guo and Pagnoni, 2008). The main limitation of TC-GICA is the assumption of the homogeneity in spatial distribution of the networks across subjects while studies have shown that functional networks can vary considerably due to subjects clinical, biological and demographic characteristics (Zhao et al., 2007; Greicius et al., 2004, among others). To address this limitation, a hierarchical ICA framework has been proposed to directly account for between-subject differences in group ICA decomposition and further allows for modeling subjects' covariate effects in ICA (Guo and Tang, 2013; Shi and Guo, 2016; Lukemire et al., 2018). All the aforementioned ICA methods are developed for cross-sectional imaging studies where subjects are only scanned once during the study.

In recent years, longitudinal studies have become increasingly popular in the neuroscience community. In such studies, brain imaging such as fMRI scans from the same individual are acquired repeatedly at multiple time points including the baseline as well as follow-up visit times. Within-subject changes in brain images across different time points provides great insights into effects and causal relationships in investigating changes in brain networks related to disease progression, treatment or neurodevelopment. By taking the advantage of using each subject as his/her own control, longitudinal studies are well-known to have the potentials to provide more reliable and significant scientific findings than cross-sectional studies. Existing longitudinal imaging analysis often focus on modeling fMRI brain activation or structural MRI volumetric measures across time (Calhoun et al., 2001b; Dettwiler et al., 2014; Lee et al., 2015). There has also been some work on longitudinal analysis of brain connectivity, which mainly involve modeling pairwise connectivity measures or network summary measures from a per-specified network structure (Dai et al., 2017; Wu et al., 2013; Li et al., 2009). However, methods are lacking for conducting longitudinal ICA that jointly decompose the subjects' repeatedly measured fMRI data, extract the underlying brain functional networks and studying the longitudinal effects on brain networks.

Existing group ICA methods are not suitable for modeling repeated measured images in longitudinal studies. There are only a couple of ad-hoc strategies for longitudinal ICA decomposition. The first approach is to conduct ICA separately at each time point and then take the ICs extracted from different time points for secondary longitudinal analysis. This separate analysis approach has limited capacity to evaluate changes in functional networks across time because 1) independent components do not have a natural order, it is difficult to identify matching components across different time points, especially in resting-state fMRI. 2) ICA algorithms usually have random elements in that they may find different local minima across different runs (Himberg, Hyvärinen and Esposito, 2004). This reduces the comparability of the ICs extracted separately at each visit. Another major drawback of the approach is that it ignores within-subject correlations among repeatedly measured data, which results in considerable loss of statistical power in testing covariate effects. The second ad-hoc approach is to adopt the TC-GICA framework by stacking all subjects' repeatedly measured images into a single group data matrix and performing ICA decomposition to extract common group spatial source signals. Then, subject/visit-specific IC maps are reconstructed via post-ICA analysis such as the dual regression. The longitudinal effects are then evaluated based on the reconstructed subject/visit-specific ICs. The limitations of the TC-GICA approach are that it ignores the between-subject variability in the ICA decomposition, does not take into account the random variabilities introduced in reconstructing subject/visit-specific IC maps and does not account for within-subject correlations among repeated scans in ICA decomposition. These limitations lead to loss of accuracy and efficiency in estimating and testing covariate effects on brain networks in longitudinal studies.

In this paper, we propose a longitudinal ICA (L-ICA) model that incorporates subject-level random effects and the time-dependent covariate effects in ICA decomposition to investigate temporal changes in brain networks and their associations with subjects clinical or demographic covariates. The L-ICA is a hierarchical model where the first-level of L-ICA decomposes a subject's fMRI data obtained at a visit into a linear mixture of subject/visit-specific spatial source signals or ICs, and these ICs are then modeled at the second-level of L-ICA in terms of population-level baseline source signals, visit effects, covariate effects, subject-specific random effects and subject/visit-specific random variability. To the best of our knowledge, L-ICA is the first model-based extension of ICA for longitudinal imaging analysis. L-ICA is able to account for within-subject correlations among repeated scans, provide more accurate estimates of changes in brain functional networks on both the population- and individual-level, and increase statistical power in detecting covariate effects on networks. Furthermore, L-ICA provides model-based prediction for changes in brain networks related to disease progression, treatment or neurodevelopment.

For model estimation, we proposed an exact EM algorithm which is fully traceable and simultaneously provides the estimation on population-level spatial maps and subject/visit-specific ICs. Furthermore, we propose a subspace-based approximate EM algorithm to provide more efficient computation. Results from the simulation studies and real data analysis show that the approximate EM algorithm significantly reduces the computation time while maintaining high estimation accuracy comparable to the exact EM. Moreover, we develop a statistical inference procedure for testing covariate effects in L-ICA, which demonstrates lower type I error and higher statistical power than the existing testing method based on TC-GICA. We apply the L-ICA method to investigating changes in functional networks in ADNI2 longitudinal rs-fMRI study. Results from L-ICA showed differential temporal changing patterns between Alzheimer and control groups in relevant brain networks, which is not revealed by existing ICA methods.

This paper is organized as follows. The methodology of L-ICA is presented in the section 2 which includes the L-ICA model specification, estimation via the exact EM algorithm and the approximate EM algorithm, and the inference procedure. In the section 3, results from the simulation study are presented. Section 4 is the real data application of ANDI2 study. Conclusion and discussion are in section 5.

## 2. Methods

This section introduces the L-ICA framework, which includes the model specification, EM algorithms and the inference procedure. To set the notation, suppose that in a longitudinal fMRI study, there are $N$ subjects and each of them has $K$ visits during the study. At each visit, a series of $T$ fMRI scans are acquired where each scan represents a 3D brain image containing $V$ voxels. Let $\widetilde{\boldsymbol{Y}}_{ij} = [\widetilde{\boldsymbol{y}}_{ij}(1), ..., \widetilde{\boldsymbol{y}}_{ij}(V)]$ be the $T \times V$ fMRI data matrix for subject $i$ $(i = 1, \ldots, N)$ at visit $j$ $(j = 1, \ldots, K)$ where $\widetilde{\boldsymbol{y}}_{ij}(v) \in \mathbb{R}^T$ represents the centered blood-oxygen-level dependent (BOLD) signal series at voxel $v$ ($v = 1, \ldots, V$). Prior to ICA, some preprocessing steps such as centering, dimension reduction and whitening of the observed data are usually performed to facilitate the subsequent ICA decomposition (Hyvärinen, Karhunen and Oja, 2001). Following a PPCA-based preprocessing procedure similar to that used in previous work (Beckmann and Smith, 2004; Shi and Guo, 2016; Guo and Tang, 2013), we perform the dimension reduction and whitening procedure on $\widetilde{\boldsymbol{Y}}_{ij}$ to obtain a $q \times V$ preprocessed data matrix $\boldsymbol{Y}_{ij}$ for subject $i$ at visit $j$, where $q$ is the number of independent components. More details about pre-processing are in Appendix. Throughout the rest of our paper, we will present the L-ICA model and methodologies based on the preprocessed data.

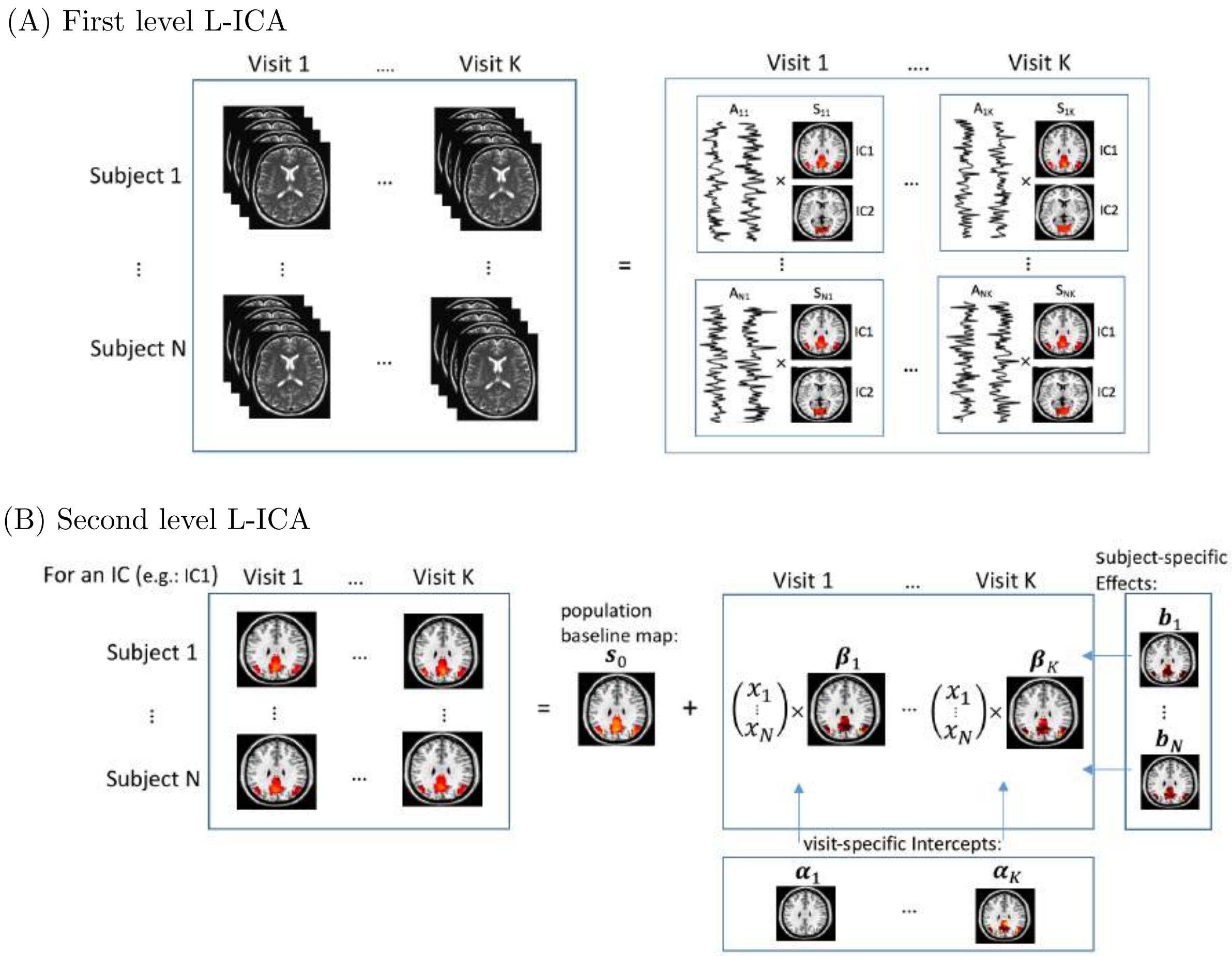

Fig 1. *Schematic illustration of the hierarchical modeling framework of L-ICA. (A) the first level model of L-ICA with $N$ subjects and $K$ visits where each subject/visit-specific fMRI data is decomposed into $q$ subject/visit-specific ICs, here $q = 2$ for illustration purpose. (B) the second level model of L-ICA for one specific IC where the subject/visit-specific ICs are modelled in terms of population-level source signals, subject specific random effects, visit effects and visit-specific covariate effects.*

2.1. *Longitudinal ICA model (L-ICA)* In this section, we propose a longitudinal ICA (L-ICA) model to jointly decompose repeated measured fMRI data acquired across multiple visits. The L-ICA is developed under a hierarchical modeling framework. We present a schematic illustration of the L-ICA in Figure 1. The first level of L-ICA decomposes the subject/visit-specific fMRI data into a product of subject/visit-specific spatial source signals and temporal mixing matrix. This allows capturing the variabilities of the functional networks across subjects and across visits. We also include a noise term in the first level ICA model to account for residual variabilities in the fMRI data that are not explained by the extracted ICs, which is known as probabilistic ICA (Beckmann and Smith, 2004). Specifically, the first level of L-ICA is as follows,

$$\text{Level 1: } \boldsymbol{y}_{ij}(v) = \boldsymbol{A}_{ij}\boldsymbol{s}_{ij}(v) + \boldsymbol{e}_{ij}(v), \tag{1}$$

where $\boldsymbol{s}_{ij}(v) = [s_{ij}^{(1)}(v), ..., s_{ij}^{(q)}(v)]'$ is a $q \times 1$ vector with $s_{ij}^{(\ell)}(v)$ $(\ell = 1, \dots, q)$ representing the spatial source signal of the $\ell$th IC (i.e., brain functional network) at voxel $v$ for subject $i$ at visit $j$, $\boldsymbol{A}_{ij}$ is the $q \times q$ mixing matrix for subject $i$ at visit $j$, which is commonly assumed to be orthogonal given that $\boldsymbol{y}_{ij}(v)$ is whitened (Hyvärinen and Oja, 2000). $\boldsymbol{e}_{ij}(v)$ is a $q \times 1$ vector that represents the noise in the subject's data and $\boldsymbol{e}_{ij}(v) \sim \mathrm{N}(\boldsymbol{0}, \boldsymbol{E}_v)$ for $v = 1, ..., V$. Prior to ICA, preliminary analysis such as pre-whitening (Bullmore et al., 1996) can be performed to remove temporal correlations in the noise term and to standardize the variability across voxels. Therefore, following previous work (Hyvärinen, Karhunen and Oja, 2001; Beckmann and Smith, 2004, 2005; Guo and Pagnoni, 2008; Guo, 2011), we assume that the covariance for the noise term is isotropic across voxels, i.e. $\boldsymbol{E}_v = \sigma_0^2 \boldsymbol{I}_q$.

At the second-level of L-ICA, we further model subject/visit-specific spatial source signals $\boldsymbol{s}_{ij}(v)$ as a combination of the population-level source signals, subject-specific random effects, visit-specific covariate effects and subject/visit-specific random variations. That is,

$$\text{Level 2: } \boldsymbol{s}_{ij}(v) = \boldsymbol{s}_0(v) + \boldsymbol{b}_i(v) + \boldsymbol{\alpha}_j(v) + \boldsymbol{\beta}_j(v)'\boldsymbol{x}_i + \boldsymbol{\gamma}_{ij}(v), \tag{2}$$

where $\boldsymbol{s}_0(v) = [s_{01}(v), ..., s_{0q}(v)]'$ is the population-level spatial source signals. The $q$ elements of $\boldsymbol{s}_0(v)$ are assumed to be independent and non-Gaussian. $\boldsymbol{b}_i(v)$ is the $q \times 1$ subject-specific random effects for $q$ ICs where $\boldsymbol{b}_i(v) \sim N(0, \boldsymbol{D})$ with $\boldsymbol{D} = \mathrm{diag}(\nu_1^2, ..., \nu_q^2)$. The subject-specific random effects help capture the within-subject correlations among the scans repeated acquired on the same subject at different visits. $\boldsymbol{\alpha}_j(v)$ is a $q \times 1$ visit effects parameter representing the population-level changes in spatial source signals from baseline to the $j$th visit. $\boldsymbol{x}_i = [x_{i1}, ..., x_{ip}]'$ is the $p \times 1$ subject-specific covariate vector which may contain a subject's clinical and demographic information such as disease group, gender, age, etc. $\boldsymbol{\beta}_j(v)$ is a $p \times q$ parameters matrix reflecting how subjects' covariates $\boldsymbol{x}_i$ modulate the subject/visit-specific brain networks. Finally, $\boldsymbol{\gamma}_{ij}(v)$ is a $q \times 1$ zero-mean Gaussian random vector, i.e. $\boldsymbol{\gamma}_i(v) \overset{\text{iid}}{\sim} \text{N}(\mathbf{0}, \tau^2 \boldsymbol{I}_q)$, capturing the residual random variability among subject/visit-specific brain networks after adjusting for the other effects in the model. In the Level 2 model, by including the subject-specific random effects, L-ICA is able to borrow information among the multiple visits within the same subject to obtain more accurate estimate of unique patterns in brain networks specific to the individual. L-ICA incorporates the visit-specific covariate effects to allow flexibly modeling time-varying covariate effects on subjects' brain networks in a longitudinal study .

2.2. *Source signal distribution model* We specify mixtures of Gaussians (MoG) as our source distribution model for the population-level spatial source signals, $\boldsymbol{s}_0(v)$. MoG has been selected as the distribution for independent components in quite a few ICA analysis (Attias, 2000; Guo, 2011; Guo and Tang, 2013; Shi and Guo, 2016) because it has several desirable properties for modeling fMRI signals. Within each brain functional network, only a small percentage of locations in the brain are activated or deactivated whereas most brain areas exhibit background fluctuations (Biswal and Ulmer, 1999). MoG are well suited to model such mixed patterns. Furthermore, MoG can capture various types of non-Gaussian signals (Xu et al., 1997; Kostantinos, 2000) and also offer tractable likelihood-based estimation (McLachlan and Peel, 2004).

Specifically, for $\ell = 1, \ldots, q$ we assume that the spatial source signal $s_{0\ell}(v)$ follows a MoG distribution, i.e.

$$s_{0\ell}(v) \sim \text{MoG}(\boldsymbol{\pi}_\ell, \boldsymbol{\mu}_\ell, \boldsymbol{\sigma}_\ell^2), \tag{3}$$

where $\boldsymbol{\pi}_\ell = [\pi_{\ell,1}, ..., \pi_{\ell,m}]'$ with $\sum_{j=1}^m \pi_{\ell,j} = 1$ is the weight parameters in MoG, $\boldsymbol{\mu}_\ell = [\mu_{\ell,1}, ..., \mu_{\ell,m}]'$ and $\boldsymbol{\sigma}_\ell^2 = [\sigma_{\ell,1}^2, ..., \sigma_{\ell,m}^2]'$ are the mean and variance parameters of the Gaussian component distributions in the MoG; $m$ is the number of Gaussian components in MoG. The probability density function of $\text{MoG}(\boldsymbol{\pi}_\ell, \boldsymbol{\mu}_\ell, \boldsymbol{\sigma}_\ell^2)$ is $\sum_{j=1}^m \pi_{\ell,j} g(s_{0\ell}(v); \mu_{\ell,j}, \sigma_{\ell,j}^2)$ where $g(\cdot)$ is the pdf of the Gaussian distribution. In fMRI applications, mixtures of two to three Gaussian components can be used to capture the distribution of fMRI spatial signals, with the different Gaussian components representing the background fluctuation and the negative or positive fMRI BOLD effects respectively (Beckmann and Smith, 2004; Guo and Pagnoni, 2008; Guo, 2011; Wang et al., 2013; Guo and Tang, 2013). Without loss of generality, we denote the first Gaussian component, i.e. $j = 1$, to be the background fluctuation state throughout the rest of the paper. To facilitate derivations with the MoG model, we introduce a voxel-specific latent state variable $z_\ell(v)$ which represents which Gaussian component in MoG that voxel $v$ belongs to. Specifically, $z_\ell(v)$ takes a value in $\{1, \ldots, m\}$ with probability $p[z_\ell(v) = j] = \pi_{\ell,j}$ $(j = 1, .., m)$. When $z_\ell(v) = j$, the $v$th voxel follows the $j$th Gaussian component distribution in MoG, i.e. $p(s_{0\ell}(v)|z_\ell(v) = j) = g(s_{0\ell}(v); \mu_{\ell,j}, \sigma_{\ell,j}^2)$.

2.3. *Maximum likelihood estimation and the EM algorithm* The parameters in L-ICA model is estimated via maximum likelihood (ML) approach. Based on the hierarchical models in (1) and assuming the independence among voxels, (2) and (3), the complete data log-likelihood for L-ICA model is,

$$l(\Theta; \mathcal{Y}, \mathcal{X}, \mathcal{S}, \mathcal{B}, \mathcal{Z}) = \sum_{v=1}^{V} l_v(\Theta; \mathcal{Y}, \mathcal{X}, \mathcal{S}, \mathcal{B}, \mathcal{Z}), \tag{4}$$

where $\mathcal{Y} = \{\boldsymbol{y}_{ij}(v) : i = 1, ..., N; j = 1, ..., K; v = 1, \ldots, V\}$ are the preprocessed longitudinal fMRI data across subjects, $\mathcal{X} = \{\boldsymbol{x}_i : i = 1, ..., N\}$ are subjects' covariates, $\mathcal{S} = \{\boldsymbol{s}_0(v), \boldsymbol{s}_{ij}(v) : i = 1, ..., N; j = 1, ..., T; v = 1, ..., V\}$ are the latent independent component spatial source signals, $\mathcal{B} = \{\boldsymbol{b}_i(v) : i = 1, ..., N; v = 1, ..., V\}$ are the subject-specific random effects and $\mathcal{Z} = \{\boldsymbol{z}(v) : v = 1, ..., V\}$ are the latent states for MoG source distribution model; the parameters in L-ICA are denoted by $\Theta = \{\{\boldsymbol{\alpha}_j(v)\}, \{\boldsymbol{\beta}_j(v)\}, \{\boldsymbol{A}_{ij}\}, \boldsymbol{E}, \boldsymbol{D}, \tau, \{\boldsymbol{\pi}_\ell\}, \{\boldsymbol{\mu}_\ell\}, \{\boldsymbol{\sigma}^2_\ell\} : i = 1, ..., N, j = 1, ..., K,$ $v = 1, ..., V, \ell = 1, ..., m\}$.

Since our likelihood function involves unobserved latent variables, we consider the expectation-maximization (EM) framework (Dempster, Laird and Rubin, 1977) for finding the maximum likelihood estimates of parameters. The EM algorithm is an iterative algorithm that alternates between performing an expectation step (E-step) and a maximization step (M-step). In the E-step, we compute an expectation of the log-likelihood conditioning on the distribution of latent variables given the observed data $\mathcal{Y}$ and the current parameter estimates $\hat{\Theta}^{(k)}$. At the M-step, the updated maximum likelihood estimates of the parameters is computed by maximizing the expected log-likelihood found on the E-step. The parameter estimates found on the M-step are then used to begin another E-step, and the process is iterated until convergence, i.e. until the parameter estimates $\hat{\Theta}^{(k)}$ and $\hat{\Theta}^{(k+1)}$ in two consecutive iterations are considered sufficiently close. In the following, we present two EM algorithms for solving the L-ICA model. The first is an exact EM method that provides exact evaluation of the conditional expectation in the E-step. We then propose an approximation EM algorithm is computationally more efficient especially with large number of ICs.

2.3.1. *The exact EM algorithm* We first develop an exact EM which has an explicit E-step and M-step to obtain ML estimates for the parameters in L-ICA.

**E-step**: In the E-step, given the estimated parameter $\hat{\Theta}^{(k)}$ from the last step, we evaluate the conditional expectation of the complete data log-likelihood as follows,

$$Q(\Theta|\hat{\Theta}^{(k)}) = \sum_{v=1}^{V} E_{\boldsymbol{L}(v)|\mathbf{y}(v), \hat{\Theta}^{(k)}} \left[ l(\Theta; \mathcal{Y}, \mathcal{X}, \mathcal{S}, \mathcal{B}, \mathcal{Z}) \right], \tag{5}$$

where $\boldsymbol{L}(v) = [\boldsymbol{b}_1(v)', ..., \boldsymbol{b}_N(v)', \boldsymbol{s}_0(v)', \boldsymbol{s}_{11}(v)', ..., \boldsymbol{s}_{NK}(v)']'$ are the latent variables in L-ICA model which include the latent source signals on both the population and individual level and the subject-specific random effects. To calculate the conditional expectation, we need to derive the conditional distribution of $\boldsymbol{L}(v)$ given the observed data $\mathbf{y}(v)$, i.e. $p(\boldsymbol{L}(v) \mid \mathbf{y}(v), \hat{\Theta}^{(k)})$. To facilitate this derivation, we take the following steps. First, we derive the distribution of $\boldsymbol{L}(v)$ given both the observed data $\mathbf{y}(v)$ and the latent states $\boldsymbol{z}(v)$, i.e. $p(\boldsymbol{L}(v) \mid \mathbf{y}(v), \boldsymbol{z}(v), \hat{\Theta}^{(k)})$, which can be shown to be a multivariate Gaussian distribution. Next, we derive the conditional distribution of the latent states given the observed data, i.e. $p[\boldsymbol{z}(v) \mid \mathbf{y}(v), \hat{\Theta}^{(k)}]$, by applying the Bayes' Theorem. Finally, we obtain the conditional distribution of $\boldsymbol{L}(v)$ given $\mathbf{y}(v)$ by integrating out $\boldsymbol{z}(v)$, i.e.

$$p(\boldsymbol{L}(v) \mid \mathbf{y}(v), \hat{\Theta}^{(k)}) = \sum_{\boldsymbol{z}(v)\in\mathcal{R}} p(\boldsymbol{L}(v) \mid \mathbf{y}(v), \boldsymbol{z}(v), \hat{\Theta}^{(k)}) p[\boldsymbol{z}(v) \mid \mathbf{y}(v); \hat{\Theta}^{(k)}],$$

where $\mathcal{R}$ represents the set of all possible values of $\boldsymbol{z}(v)$, i.e., $\mathcal{R} = \{\boldsymbol{z}^r\}_{r=1}^{m^q}$ where $\boldsymbol{z}^r = [z_1^r, ..., z_q^r]'$ and $z_\ell^r \in \{1, ..., m\}$ for $\ell = 1, ...q$.

Following this procedure, we can derive explicit form for the conditional distribution for the latent variables and subsequently deriving the conditional expectation $Q(\Theta|\hat{\Theta}^{(k)})$ in (5).

**M-step**: In the M-step, the updated estimates are obtained by maximizing the expected log-likelihood function computed in the E-step, i.e.,

$$\hat{\Theta}^{(k+1)} = \underset{\Theta}{\operatorname{argmax}}\, Q(\Theta|\hat{\Theta}^{(k)}). \tag{6}$$

We have derived explicit solutions for all parameter updates (please see Appendix for details).

The steps of the exact EM algorithm is summarized in Algorithm 1. The detailed derivations are presented in the Appendix.

After obtaining the ML estimates $\hat{\Theta}$, we estimate the baseline population- and subject/visit-specific source signals as well as their variability based on the mean and variance of their conditional distributions, i.e., $[\boldsymbol{s}_0(v) \mid \mathbf{y}(v); \hat{\Theta}]$ and $[\boldsymbol{s}_{ij}(v) \mid \mathbf{y}(v); \hat{\Theta}]$. These conditional moments are directly obtainable from the E-step of our algorithm upon convergence and no separate post-ICA steps are required. Based on the estimated covariate effects $\{\hat{\boldsymbol{\beta}}(v)\}$, we can investigate how subjects' clinical and demographic characteristics affects their brain functional networks and their changes across visits. Furthermore, the L-ICA also provides model-based prediction of the brain functional networks for specific sub-populations at a given visit. For example, for a sub-population characterized by a covariates pattern $\boldsymbol{x}^*$, the predicted brain functional networks at the $j$th visit can be derived by plugging the ML parameter estimates into Level 2 of L-ICA, i.e.

$$\hat{\boldsymbol{s}}_j(v) = \hat{\boldsymbol{s}}_0(v) + \hat{\boldsymbol{\alpha}}_j(v) + \hat{\boldsymbol{\beta}}_j(v)'\boldsymbol{x}^*, \tag{7}$$

**Algorithm 1** The Exact EM Algorithm

**Initial values**: Obtain an initial values $\hat{\Theta}^{(0)}$ based on existing group ICA software.
**repeat**
**E-step:**
1. Evaluate the conditional distribution of the latent variables $p(\boldsymbol{L}(v) \mid \mathbf{y}(v), \hat{\Theta}^{(k)})$ using the proposed three-step approach:
1.a Evaluate the multivariate Gaussian $p[\boldsymbol{L}(v) \mid \mathbf{y}(v), \boldsymbol{z}(v), \hat{\Theta}^{(k)}]$;
1.b Evaluate $p[\boldsymbol{z}(v) \mid \mathbf{y}(v); \hat{\Theta}^{(k)}]$ via Bayes' Theorem
1.c integrate out the latent states $\boldsymbol{z}(v)$

$$p(\boldsymbol{L}(v) \mid \mathbf{y}(v), \hat{\Theta}^{(k)}) = \sum_{\boldsymbol{z}(v) \in \mathcal{R}} p(\boldsymbol{L}(v) \mid \mathbf{y}(v), \boldsymbol{z}(v), \hat{\Theta}^{(k)}) p[\boldsymbol{z}(v) \mid \mathbf{y}(v); \hat{\Theta}^{(k)}]$$

2. Estimate conditional expectation $Q(\Theta|\hat{\Theta}^{(k)})$ based on $p(\boldsymbol{L}(v) \mid \mathbf{y}(v), \hat{\Theta}^{(k)})$.
**M-step:**
Update parameters estimates

$$\hat{\Theta}^{(k+1)} = \underset{\Theta}{\operatorname{argmax}}\, Q(\Theta|\hat{\Theta}^{(k)}).$$

**until convergence**, i.e. $\frac{\|\hat{\Theta}^{(k+1)} - \hat{\Theta}^{(k)}\|}{\|\hat{\Theta}^{(k)}\|} < \epsilon$

2.4. *Subspace EM algorithms* The exact EM algorithm requires $\mathcal{O}(m^q)$ operations at each voxel which is an exponential increase with regard to the number of the ICs extracted in L-ICA, which will be time consuming when $q$ is large. The reason for needing $\mathcal{O}(m^q)$ operations is that, the exact EM evaluates the conditional distribution of the latent states $\boldsymbol{z}(v)$, i.e. $p[\boldsymbol{z}(v) \mid \mathbf{y}(v)]$, across the whole sample space $\mathcal{R}$ of $\boldsymbol{z}(v)$, which has a cardinality of $m^q$. To reduce the computation load, we develop a subspace-based approximate EM for L-ICA model. The motivation of the subspace EM is based on the observation from fMRI analysis that the density of $p[\boldsymbol{z}(v) \mid \mathbf{y}(v)]$ is mostly concentrated on a subspace $\mathcal{R}_s = \{\boldsymbol{z}^r \in \mathcal{R},\ s.t. \sum_\ell I(z^r_\ell \neq 1) \leq 1\}$. To help understand this subspace, recall that the latent state $z^r_\ell$ takes values in $(1, \ldots, m)$ with the first state, i.e. $z^r_\ell = 1$, corresponding to the background fluctuation while other states, i.e. $z^r_\ell \neq 1$, corresponding to either positive or negative signals at a voxel. Therefore, the subspace $\mathcal{R}_s$ corresponds to that a voxel has active signals in at most one of the $q$ ICs. In Shi and Guo (2016), we have provided theoretical proof that the density of the conditional distribution of the latent states is mostly concentrated in the subspace $\mathcal{R}_s$ when the source signals are sparse in each IC, which is the case with the fMRI spatial source signals which have been shown to be sparse across the brain (Mckeown et al., 1998; Daubechies et al., 2009).

In the subspace EM algorithm, we follow the similar steps as in the exact EM algorithm presented in Algorithm 1. The main difference is that when evaluating and summing across the latent states $\boldsymbol{z}(v)$ in the E-step and M-step, we replace the whole sample space $\mathcal{R}$ with the proposed subspace $\mathcal{R}_s$ which only has carnality of $(m-1)q+1$. This means the subspace EM only requires $\mathcal{O}(mq)$ operations at each voxel which scales linearly with the number of ICs and is significantly faster than the exponential growth of the exact EM algorithm.

2.5. *Statistical inference for testing covariate effects in L-ICA* In this session, we propose a statistical inference procedure for testing covariate effects in L-ICA to investigate whether the covariates have significant effects on brain functional networks and their changes across visits. Typically, statistical inference in maximum likelihood estimation is conducted by inverting the information matrix to estimate the variance-covariance matrix of ML estimates of the parameters. However, this standard approach is not feasible when modeling fMRI data with L-ICA because the high dimensionality of the parameter space makes extremely challenge to obtain a reliable inversion of information matrix. To address this issue, we develop a computational efficient statistical inference procedure based on the connection between the L-ICA and multivariate linear models. The proposed inference procedure provides an efficient approach to estimate the variance-covariance matrix of the time-specific covariate effects at each voxel by directly using the output from our EM algorithms.

Specifically, let $\boldsymbol{y}_i(v)$ be the $i$th subjects longitudinal fMRI data which is a $qK \times 1$ vector obtained by stacking his/her data across visits, i.e. $\boldsymbol{y}_i(v) = [\boldsymbol{y}_{i1}(v)', ..., \boldsymbol{y}_{iK}(v)']'$. By collapsing the hierarchical models, we rewrite the L-ICA model in a non-hierarchical form which is similar to classical multivariate linear model, i.e.,

$$\boldsymbol{y}_i^*(v) = \boldsymbol{X}_i^* \boldsymbol{C}^*(v) + \boldsymbol{\zeta}_i(v), \tag{8}$$

where $\boldsymbol{y}_i^*(v) = \boldsymbol{A}_i' \boldsymbol{y}_i(v)$ is the response vector, $\boldsymbol{X}_i^*$ is the design matrix which includes the visit time and the covariates in L-ICA, $\boldsymbol{C}^*(v)$ is the parameter matrix which includes the effects parameters in L-ICA such as the visit effects $\boldsymbol{\alpha}$ and covariate effects $\boldsymbol{\beta}$, $\boldsymbol{\zeta}_i(v)$ is the zero-mean Gaussian random variation term which includes the subject-specific random effects and noise terms in L-ICA. Please see the Appendix for details.

The model in (8) can be viewed a multivariate linear model. Based on linear model theory, a variance estimator for parameter estimates $\hat{\boldsymbol{C}}^*(v)$ can be derived as follows,

$$\mathrm{Var}[\hat{\boldsymbol{C}}^*(v)] = \left( \sum_{i=1}^{N} \boldsymbol{X}_i^{*\prime} \boldsymbol{W}(v)^{-1} \boldsymbol{X}_i^* \right)^{-1}. \tag{9}$$

where $\boldsymbol{W}(v) = \mathrm{Var}(\boldsymbol{\zeta}_i(v))$ and can be estimated by plugging ML estimates obtained from the EM algorithm.

After deriving the variance estimator for the ML estimates of the parameters in L-ICA, We can then conduct hypothesis testing on the covariate effects the brain networks and their changes across visits. Specifically, we first formulate the hypothesis in terms of linear combinations of the parameters in the L-ICA model, i.e. $H_0 : \boldsymbol{l}'\boldsymbol{C}^*(v) = \mathbf{0}$ vs. $H_1 : \boldsymbol{l}'\boldsymbol{C}^*(v) \neq \mathbf{0}$ where $\boldsymbol{l}$ is a vector of constant coefficients specified based on the hypothesis that we are testing on. We can then construct the test statistic as,

$$z(v) = \frac{\boldsymbol{l}'\hat{\boldsymbol{C}}^*(v)}{\sqrt{\boldsymbol{l}'\hat{\text{Var}}[\hat{\boldsymbol{C}}^*(v)]\boldsymbol{l}}}, \tag{10}$$

the test statistic $z(v)$ will then be compared against its null distribution to derive the p-value for testing the significance of the covariate effects at voxel $v$. Standard multiple testing correction procedures can be applied to control for family wise error rate (FWER) or the false discovery rate (FDR) when testing the covariate effects across voxels, (Genovese, Lazar and Nichols, 2002; Chumbley and Friston, 2009; Storey, 2011; Wang, Wu and Yu, 2017).

**3. Simulation Study** We conducted three types of simulation studies to 1) evaluate the performance of the proposed L-ICA model as compared with the approach based on the existing TC-GICA framework, 2) to evaluate the performance of the proposed inference method for testing covariate effects on brain networks, and 3) to evaluate the performance of the proposed subspace-based EM algorithm as compared with the exact EM algorithm.

3.1. *Simulation study I: performance of the L-ICA v.s. TC-GICA-based longitudinal analysis* In this simulation study, we evaluate the performance of the proposed L-ICA model versus the TC-GICA based approach for analyzing longitudinal fMRI. In the simulation, we considered three different sample sizes $N = 10, 20, 60$ and each subject has three visits: baseline, visit 1 and visit 2 ($K = 3$). The simulated fMRI data were generated from 3 underlying ICs or source signals, i.e., $q = 3$, (see Figure 2 (A)). For each IC, we generated the source signals $\{\boldsymbol{s}_0(v)\}$ as a 3D spatial map with the dimension of $53 \times 63 \times 3$, which was based on three selected slices from a real fMRI imaging data. The source intensity at the activated region in the IC maps was generated from a Gaussian distribution with the mean of 4. The visit specific intercepts, i.e., $\boldsymbol{\alpha}_2(v)$ and $\boldsymbol{\alpha}_3(v)$, are set to be 2 and 3 respectively for the voxels within the activated IC regions and 0 for other voxels. We then generated a binary covariate for each subject as $x_i \overset{\text{iid}}{\sim} \text{Bernoulli}(0.5)$. The covariate effects at the $j$th visit, $\boldsymbol{\beta}_j(v)$, were specified using a 2D Gaussian process within the IC regions where the mean level of the covariate effects increased across the 3 visits. Additionally, we generated subject-specific random effects, i.e., $\boldsymbol{b}_i(v)$, from a zero-mean Gaussian distribution with the covariance matrix of $\boldsymbol{D} = \text{diag}(1.0^2, 1.1^2, 1.2^2)$. For the residual subject/visit-specific variability, i.e., $\boldsymbol{\gamma}_i(v)$, we considered two levels of variability: low ($\tau^2 = 0.5$) and high ($\tau^2 = 4$). The time series associated with each IC was generated from real fMRI time courses with the length of $T = 200$ and hence represented realistic fMRI temporal dynamics. We generated subject/visit-specific time sources that had similar frequency features but different phase patterns (Guo, 2011; Shi and Guo, 2016), which mimic temporal dynamics in resting-state fMRI. After simulating the spatial source signal and the temporal mixing matrices for the ICs, Gaussian background noise with a standard deviation of 1 (i.e. $\boldsymbol{E} = \boldsymbol{I}_q$) were added to generate observed fMRI data.

Following previous work (Beckmann and Smith, 2005; Guo and Pagnoni, 2008; Guo, 2011), we evaluate the performance of each method based on the correlations between the true ICs and estimated ICs in both temporal and spatial domains. We report the estimation accuracy for both the population-level as well as the subject/visit-specific source signals. To compare the performance in estimating the covariate effects, we report the mean square errors (MSEs) of $\hat{\boldsymbol{\beta}}(v)$ defined by $\frac{1}{KV}\sum_{j=1}^{K}\sum_{v=1}^{V}\left\|\hat{\boldsymbol{\beta}}_j(v) - \boldsymbol{\beta}_j(v)\right\|_{\mathcal{F}}^2$ averaged across simulation runs. Here $\|\cdot\|_{\mathcal{F}}$ is the Frobenius norm for a matrix. Since ICA recovery is permutation invariant, the estimated ICs were matched to the true IC with which it has the highest spatial correlation. We present the simulation results in Table 1. The results show that L-ICA provides more accurate estimates for the source signals on both the population- and subject/visit-level, by demonstrating higher correlation with the true source signals. L-ICA also provides more accurate estimation of the covariate effects with smaller mean square errors (MSE). Moreover, compared with the TC-GICA, the L-ICA estimates of the source signals and covariate effects are more stable with consistently smaller standard deviations (SD) across simulation runs.

We also display the estimated population-level IC maps at baseline and the last visit , i.e. visit 2, based on both methods in Figure 2. The L-ICA shows better accuracy in recovering the true activation patterns in the ICs at both visits. The intensity of the source signals in the activated regions in each IC increases from baseline to the last visit in true IC maps. This increase in intensity is well captured by the L-ICA estimated IC maps but not obvious in the TC-GICA estimated IC maps. Furthermore, the estimated IC maps from the TC-GICA approach show "cross-talk" between the ICs. In Figure 2, we also present the true and estimated longitudinal trends of source signals for activated voxels in an IC. The L-ICA shows better performance than the TC-GICA approach in recovering the temporal changing patterns across voxels.

3.2. *Simulation study II: performance of the proposed inference procedure for testing covariate effects* In this simulation study, we evaluate the performance of the methods in testing covariates effects on ICs. We simulated fMRI datasets with two source signals ($q = 2$), two visits ($K = 2$), one binary covariate and the sample size of $N = 40$. Since we need a large number of simulation runs to estimate the type I error and power in the test, we generated source signal images with the dimension

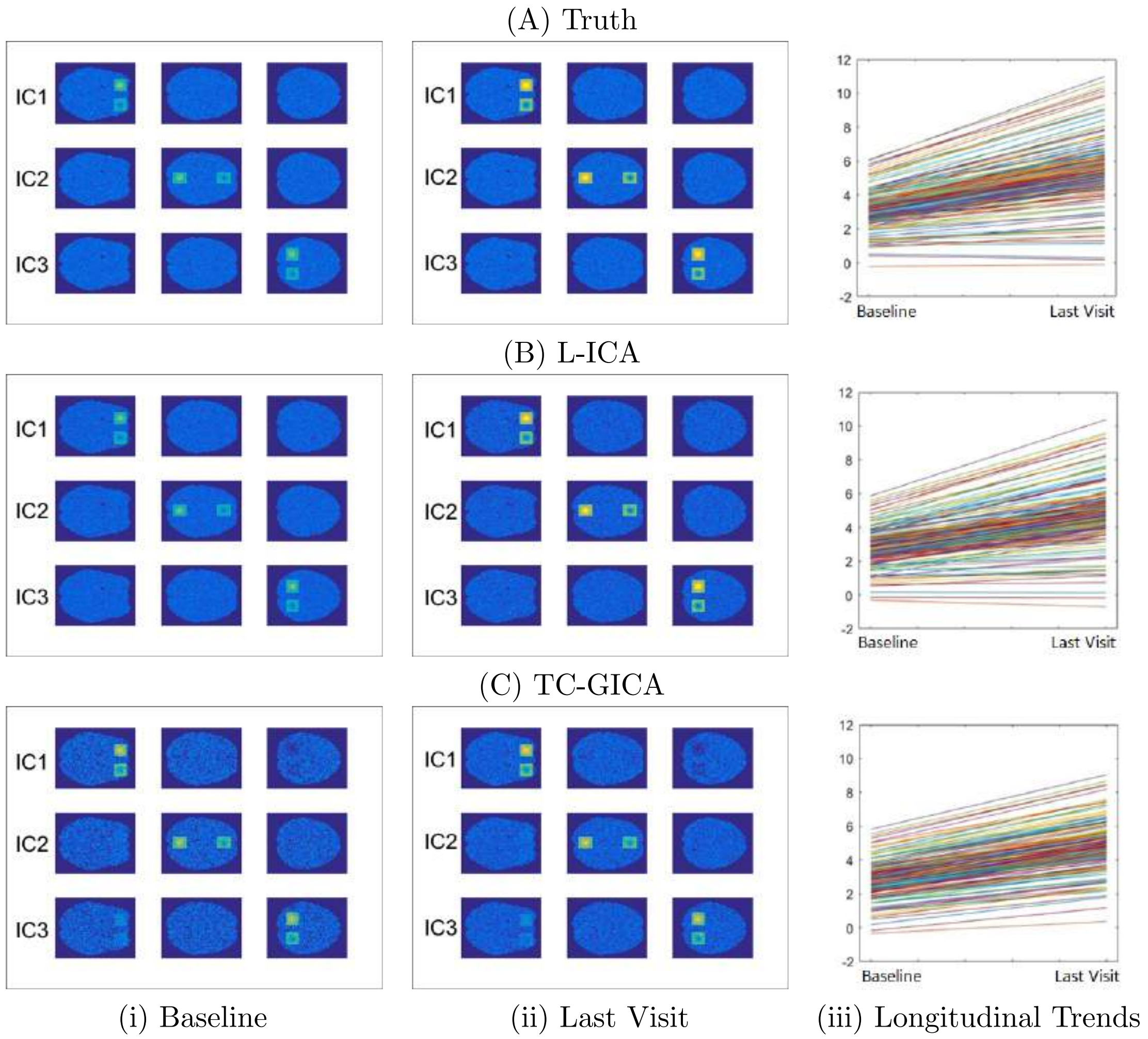

Fig 2. *Comparison between the proposed L-ICA and the TC-GICA based approach for estimating the population-level IC maps at baseline and the last visit (N=20, low subject/visit-specific random variability): (A) truth, (B) L-ICA estimates and (C) estimates from TC-GICA. Column (i) represents the IC maps at baseline ; Column (ii) represents the IC maps at last visit; Column (iii) represents the longitudinal trends for activated voxels (where each line represents a voxel) in the first IC (IC1). Results show that L-ICA provides more accurate estimates than TC-GICA at each visit and more precisely captures the voxel-specific longitudinal trend.*

TABLE 1

*Simulation results for comparing L-ICA method against TC-GICA-based method with 100 simulation runs. Values presented are mean and standard deviation of correlations between the true and estimated: population-level spatial maps, subject/visit-specific spatial maps and subject/visit-specific time courses. The mean and standard deviation of the MSE of the covariate effects estimates are also provided.*

| Subj-Visit Var | Population-level spatial maps Corr.(SD) | | Subject/Vist-specific spatial maps Corr.(SD) | |
|---|---|---|---|---|
| | L-ICA | TC-GICA | L-ICA | TC-GICA |
| Low | | | | |
| N=10 | 0.929 (0.021) | 0.853 (0.116) | 0.979 (0.016) | 0.942 (0.095) |
| N=20 | 0.959 (0.015) | 0.889 (0.113) | 0.981 (0.012) | 0.937 (0.093) |
| N=60 | 0.984 (0.008) | 0.940 (0.109) | 0.999 (0.007) | 0.951 (0.085) |
| High | | | | |
| N=10 | 0.886 (0.053) | 0.621 (0.213) | 0.960 (0.044) | 0.845 (0.152) |
| N=20 | 0.899 (0.042) | 0.691 (0.187) | 0.962 (0.034) | 0.854 (0.141) |
| N=60 | 0.958 (0.011) | 0.856 (0.162) | 0.991 (0.019) | 0.900 (0.099) |
| Subj-Visit Var. | Subject/Vist-specific time courses Corr.(SD) | | Covariate Effects MSE(SD) | |
| | L-ICA | TC-GICA | L-ICA | TC-GICA |
| Low | | | | |
| N=10 | 0.997 (0.004) | 0.941 (0.076) | 0.152 (0.009) | 0.159 (0.068) |
| N=20 | 0.998 (0.003) | 0.942 (0.075) | 0.093 (0.006) | 0.153 (0.063) |
| N=60 | 1.000 (0.001) | 0.957 (0.063) | 0.040 (0.000) | 0.128 (0.039) |
| High | | | | |
| N=10 | 0.987 (0.019) | 0.884 (0.092) | 0.253 (0.015) | 0.273 (0.101) |
| N=20 | 0.990 (0.014) | 0.885 (0.093) | 0.187 (0.011) | 0.239 (0.086) |
| N=60 | 0.992 (0.007) | 0.910 (0.077) | 0.098 (0.004) | 0.192 (0.083) |

of $20 \times 20$ to facilitate computation. The covariate effects at baseline $\beta_0(v)$ are set to be 0 representing no difference at baseline and visit-specific covariate effects $\beta_1(v)$ took values in $\{0, 0.375, 0.5, 0.625, 0.75, 0.875, 1, 1.125, 1.25\}$ for the IC region and are set to 0 for background region.

We applied L-ICA method and TC-GICA method to the simulated datasets and tested for covariate effects using both methods. We considered two type of hypothesis tests. The first one aims to test whether the covariate has an effect on the network source signals at a given visit, where the hypotheses are $H_0 : \beta_1(v) = 0$ versus $H_1 : \beta_1(v) \neq 0$ for the given IC. In the second test, we assess the whether the covariate's effect on the network vary across visits, or equivalently whether the covariate affect the longitudinal changes in the network across visits, where the hypotheses are $H_0 : \beta_1(v) = \beta_0(v)$ versus $H_0 : \beta_1(v) \neq \beta_0(v)$. These two type of tests are the most commonly conducted in longitudinal studies. For L-ICA, hypothesis tests were conducted using the test proposed in section 2.5. For TC-GICA based approach, covariate effects were tested by performing post-ICA longitudinal analysis of the dual-regression reconstructed subject/visit-specific IC maps. We estimated the Type-I error rate with the empirical probabilities of not rejecting $H_0$ at voxels where $H_0$ is true. We estimated the power of the tests with the empirical probabilities of rejecting $H_0$ at voxels where $H_1$ is true.

We report the Type-I error rates and the statistical power for detecting covariate effects based on 1000 simulation runs in Figure 3. The panel (A) in Figure 3 presents the Type I error rates where the diagonal line represents the nominal level for the type I error corresponding to various significance levels. The proposed L-ICA test demonstrates lower type-I error rates which are closer to the nominal level as compared with the TC-GICA method. For the power analysis presented in panel (B), the L-ICA have much higher statistical power in detecting covariate effects than the TC-GICA method. Overall, these results indicate that L-ICA provides more reliable and powerful statistical tests for assessing covariate effects on the functional networks.

3.3. *Simulation study III: performance of the subspace EM algorithm for LICA* In this section, we examined the performance of the subspace approximate EM algorithm as compared with the exact EM algorithm for the L-ICA model. We simulated data for ten subjects and considered three different number of ICs: $q = 3, 5, 10$. We summarize the results based on the two EM algorithms in Table 2. Results show that the accuracy of the subspace EM is comparable to that of the exact EM. The major advantage of the subspace EM is that it was much faster than the exact EM. This advantage becomes more clear with the increase of the number of ICs. For $q = 10$, the subspace-based EM only uses about 2% computation time of the exact EM.

## 4. Application to longitudindal rs-fMRI data from ADNI2 study

4.1. *Rs-fMRI acquisition and description* We applied the proposed L-ICA method to the longitudinal rsfMRI data from the Alzheimer's Disease Neuroimaging Initiative 2 (ADNI2) study. One of the main purposes of the ADNI2 project is to examine changes in neuroimaging with the progression of mild cognitive impairment (MCI) and Alzheimer's Disease(AD). Data used in our analysis were downloaded from ADNI website (http://www.adni.loni.usc.edu) and included longitudinal rs-fMRI images that were collected at baseline screening, 1 year and 2 year for four disease groups, i.e. Alzheimer's Disease (AD), late mild cognition impairment (LMCI), early mild cognition impairment (EMCI) and control (CN). A T1-weighted high-resolution anatomical image (MPRAGE) and a series of resting state functional images were acquired with 3.0 Tesla MRI scanner (Philips Systems) during longitudinal visits. The rs-fMRI scans were acquired with 140 volumnes, TR/TE = 3000/30 ms, flip angle of 80 and effective voxel resolution of 3.3x3.3x3.3 mm. More

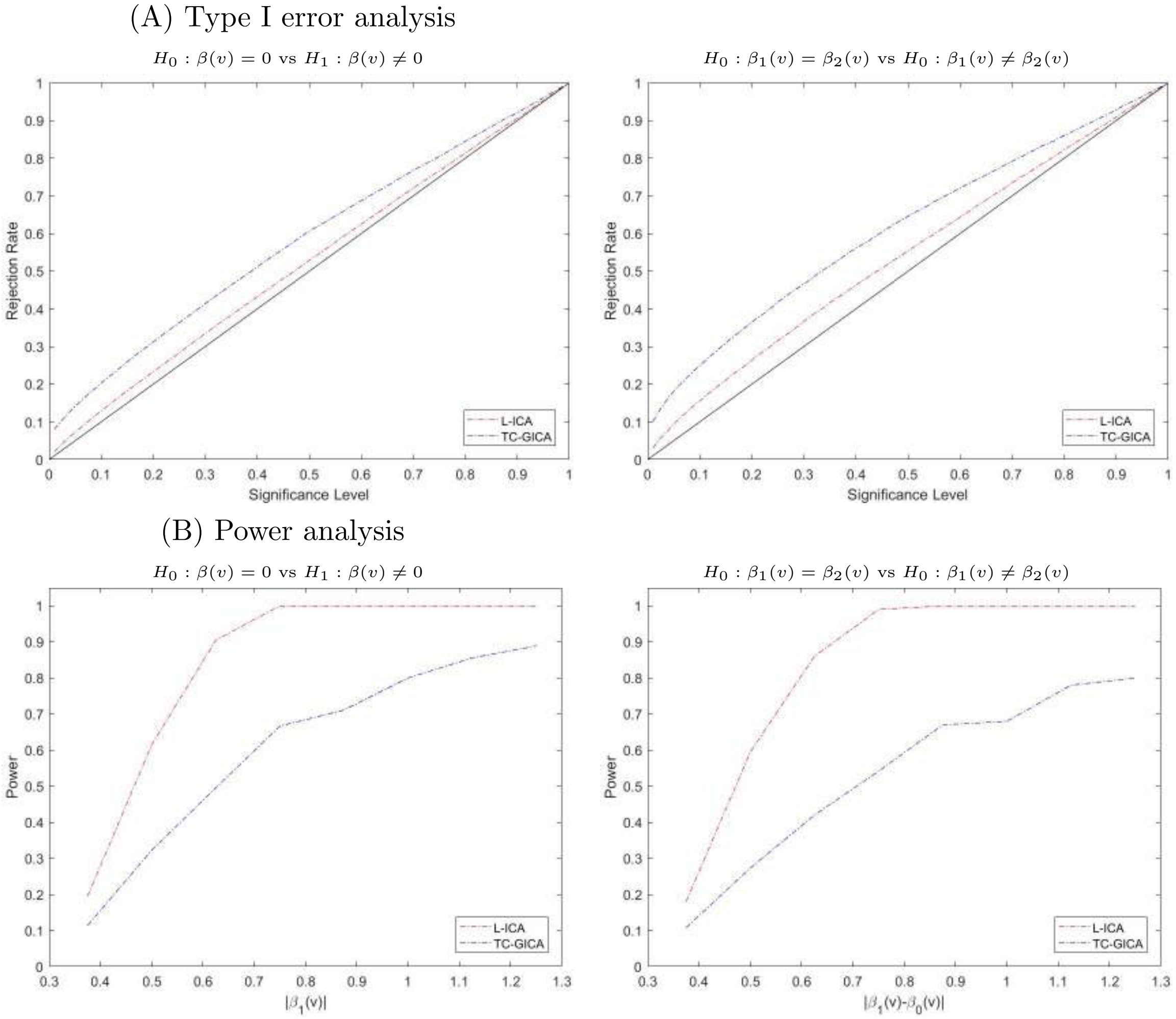

FIG 3. *Simulation results for testing covariate effects based on 1000 runs with sample size $N = 40$ using the proposed L-ICA method and the TC-GICA based method. We considered two types of hypothesis tests: testing the time-specific covariate effect at a given visit, i.e. $H_0 : \beta_1(v) = 0$ (the left column), and testing the time-varying longitudinal covariate effects, i.e. $H_0 : \beta_1(v) = \beta_2(v)$ (the right column). Panel (A) and (B) presents the type 1 error rates and the statistical power, respectively. The results show that the L-ICA method demonstrates lower type I error and higher statistical power as compared with the TC-GICA based method.*

Table 2

*Simulation results for comparing subspace EM against exact EM based on 50 simulation runs. Values presented are mean and standard deviation of the computational/iteration time (in second), the mean and standard deviation of correlations between the true and estimated: baseline population-level spatial maps and subject/visit-specific time courses, the mean and standard deviation of the MSE of the covariates estimates. The stopping criteria is based on the correlation between true and estimated subject/visit-specific spatial maps to reach 0.99 for $q = 3, 5$ and 0.90 for $q = 10$.*

| | Iteration time (SD) | | Baseline population-level spatial maps Corr.(SD) | |
|---|---|---|---|---|
| # of IC | Exact EM | Subspace EM | Exact EM | Subspace EM |
| q=3 | 98.77(2.53) | 55.26(0.85) | 0.963(0.001) | 0.962(0.001) |
| q=5 | 387.08 (5.61) | 89.42(4.51) | 0.962(0.005) | 0.961(0.004) |
| q=10 | 11254.67(9.01) | 187.82(6.31) | 0.913(0.010) | 0.907(0.009) |
| | **Subject/Visit-specific time courses Corr.(SD)** | | **Covariate Effects MSE(SD)** | |
| # of IC | Exact EM | Subspace EM | Exact EM | Subspace EM |
| q=3 | 0.998(0.003) | 0.998(0.003) | 0.083(0.009) | 0.081(0.009) |
| q=5 | 0.996(0.004) | 0.995(0.003) | 0.083(0.011) | 0.085(0.010) |
| q=10 | 0.989(0.010) | 0.986(0.007) | 0.097(0.023) | 0.102(0.021) |

details can be found at ADNI website (http://www.adni.loni.usc.edu). Quality control was performed on the fMRI images both by following the Mayo clinic quality control documentation (version 02-02-2015) and by visual examination. After the quality control, 51 subjects were included for the following ICA analysis. Among these subjects, 6 are diagnosed with AD, 17 are diagnosed with EMCI, 12 are diagnosed with LMCI and 16 are normal controls (CN) at baseline. For gender, there are 2 (33.3%) males for AD, 10 (58.8%) males for EMCI, 7 (58.3%) males for LMCI and 8 (50.0%) males for CN. The mean (SD) of age for each group is 80.3 (4.5) for AD, 72.8 (6.2) for EMCI, 70.0 (7.1) for LMCI and 74.8 (4.7) for CN. Based on F tests, there is no significant between-group difference in gender (p-value = 0.734) but significant difference in age across the groups (p-value = 0.008). We included both gender and age as covariates in the following L-ICA modeling to control for any potential confounding effects.

4.2. *Rs-fMRI preprocessing* Skull stripping was conducted on the T1 images to remove extra-cranial material. The first 4 volumes of the fMRI were removed to stablize the signal, leaving 136 volumes for subsequent prepossessing. We registered each subject's anatomical image to the 8th volume of the slice-time-corrected functional image and then the subjects' images were normalized to MNI standard brain space. Spatial smoothing with a 6mm FWHM Gaussian kernel and motion corrections were applied to the function images. A validated confound regression approach (Satterthwaite et al., 2014; Wang et al., 2016; Kemmer et al., 2015) was performed on each subject's rs-fMRI time series data to remove the potential confounding factors including motion parameters, global effects, white matter (WM) and cerebrospinal fluid (CSF) signals. Furthermore, motion-related spike regressors were included to bound the observed displacement and the functional time series data were band-pass filtered to retain frequencies between 0.01 and 0.1 Hz which is the relevant range for rs-fMRI. Lastly, we performed the prior-ICA preprocessing steps including centering, dimension reduction and whitening as described in section 2.

4.3. *L-ICA model specification for ADNI2 study* We applied the L-ICA for modeling the preprocessed baseline, 1 year and 2 year rs-fMRI data from ADNI2 study to examine the longitudinal pattern in brain networks among AD, LMCI, EMCI and CN subjects. We decomposed data into 14 ICs. The first level of L-ICA decompose subjects' longitudinal fMRI data as the product of subject/visit-specific mixing matrix and spatial source signals as specified in equation (1). In the second level model of the L-ICA, we included three binary indicators representing subjects' membership in the four disease groups (with the CN as the reference group) as our primary covariates of interest. We also included subjects' gender and baseline age as covariates to adjust for any potential confounding effects. Specifically, The second level for $l$th IC was specified as

$$s_{ij}^{(l)}(v) = s_0^{(l)}(v) + b_i^{(l)}(v) + \alpha_j^{(l)}(v) + \left(\beta_{j1}^{(l)}(v), ..., \beta_{j5}^{(l)}(v)\right)\begin{pmatrix} x_i^{AD} \\ x_i^{LMCI} \\ x_i^{EMCI} \\ x_i^{Age} \\ x_i^{Gender} \end{pmatrix} + \gamma_{ij}^{(l)}(v),$$

where $x_i^{AD} = 1$ if subject $i$ is in the AD group and 0 otherwise, and $x_i^{LMCI}$ and $x_i^{EMCI}$ are defined similarly. $\beta_{j1}^{(l)}(v)$, $\beta_{j2}^{(l)}(v)$ and $\beta_{j3}^{(l)}(v)$ represent the contrast between AD, LMCI and EMCI vs. CN, respectively, at the $j$th visit. We estimated the parameters in the L-ICA model using the subspace-based EM algorithm implemented by in-house MATLAB programs. To ensure the validity of the results from EM, we initialized the EM algorithm with 20 different initial values and the results were highly consistent.

4.4. *Longitudinal changes in brain networks for ADNI2 study based on L-ICA* Among the extracted ICs from L-ICA, we identified components that correspond to well-established brain functional networks (Smith et al., 2009) such as the default mode network (DMN), medial visual network, occipital visual network and fronto-parietal left network, which are visualized in Figures 4, 5, 6, 7. In Figure 4, we present the L-ICA model-based estimates of the DMN for the four disease groups at the three visits. The subpopulation maps were estimated at the mean baseline age (73.7 year old) and averaged between the two genders to control for confounding effects. They were thresholded based on the estimated intensity of the source signals. To provide better visualization of the changing patterns across voxels in the DMN, we also present in Figures 8 the model-based estimates of longitudinal trends of source signals for voxels in the two subregions of DMN, i.e. the posterior cingulate cortex (PCC) and the lateral parietal cortex (LPC). Figure 4 and Figure 8 shows that the four disease groups demonstrated different temporal changing patterns in the DMN source signals across the visits. Results show that the AD and LMCI patients generally have more significant changes in the DMN network across the 3 visits as compared with the EMCI and CN subjects. We also found that the longitudinal changes in the network may not necessarily follow a linear pattern and are different between the PCC and LPC regions of the DMN. Another finding from Figure 8 is that the AD group demonstrate larger variations across voxels within the network as compared with the other groups.

We also present the estimated subpopulation IC maps and the voxel-level longitudinal trends for other networks of interest (Figures 5, 6, 7, 9). Similar as the DMN, we found that the AD and LMCI patients generally have more significant changes in these networks across the 3 visits as compared with the EMCI and CN subjects, the longitudinal changes are not necessarily linear across time, and that the AD group demonstrates the larger within-network heterogeneity as compared with the other groups.

We then applied the proposed inference procedure to formally test the between-group differences at each visit while controlling for potential confounding effects from age and gender. We considered the differences between AD and CN group to demonstrate network changes in clinically diagnosed Alzheimer patients as compared with normal controls. We also considered the differences between the two MCI groups to investigate the heterogeneity between the early and late MCI stages. We then conducted tests to examine longitudinal changes from baseline to year 2 within disease group. For comparison, we applied the TC-GICA based method to examine the group differences and longitudinal differences. We illustrate results for the DMN for demonstration purpose.

Figure 10 and Figure 11 present the between-group test results for AD vs. CN and LMCI vs. EMCI, respectively. The proposed L-ICA detected significant between-group differences at each visit. Furthermore, the test results from L-ICA indicate that the between-group differences tend to increase across time with group differences observed at increasingly more spatial locations in the network. In comparison, the TC-GICA based approach identified few differences between the groups. Figure 12 represents the differences between baseline and following visits based on L-ICA. It shows that AD has more longitudinal changes compared with other groups. Specifically, Figure 13 presents the results for testing the changes from baseline screening to year 2 for AD group. Results from L-ICA show that the AD group demonstrated noticeable longitudinal changes in DMN, which are consistent with findings reported in previous work (Dai et al., 2017). In comparison, the TC-GICA approach identified very little longitudinal changes in DMN among the AD patients. As in the simula-tion studies, the results from the real data analysis show that the L-ICA method has higher statistical power in detecting group differences and longitudinal changes.

**5. Discussion** In this paper, we proposed a longitudinal ICA model (L-ICA) to formally quantify time-evolving patterns in brain function networks. In the L-ICA model, we incorporated subject-specific random effects to capture the varbilities across subjects and also borrow information across visits within the same subject to improve the model efficiency. Furthermore, to capture the possible non-linear changing effects in brain functional networks, L-ICA incorporates visit-specific covariate effects which can flexibly capture time-varying effects from subjects' demographic, clinical and biological variables. The proposed L-ICA has demonstrated lower type I error and higher statistical power in detecting covariate effects on brain networks and their changes across time.

We developed a maximum likelihood estimation method via EM algorithms for L-ICA model. Based on results from the EM, L-ICA model can simultaneously estimate population and subject/visit-specific brain functional networks. We show that L-ICA's model-based estimates of brain functional networks are more accurate on both population- and individual level. Furthermore, we proposed a computationally efficient subspace based EM algorithm. Simulation study showed that the approximate EM dramatically improves computational efficiency while achieving similar accuracy in model estimation. A Matlab toolbox will be made publicly available on the authors website for implementing the L-ICA model, (Lukemire et al., 2018).

**Acknowledgements** We thank Dr. Tian Dai for helping with downloading and preprocessing ADNI2 study data. Research reported in this publication was supported by the National Institute of Mental Health of the National Institutes of Health under Award Number R01MH105561 and R01MH079448. The content is solely the responsibility of the authors and does not necessarily represent the official views of the National Institutes of Health.

ADNI is funded by the National Institute on Aging, the National Institute of Biomedical Imaging and Bioengineering, and through generous contributions from the following: AbbVie, Alzheimer's Association; Alzheimer's Drug Discovery Foundation; Araclon Biotech; BioClinica, Inc.; Biogen; Bristol-Myers Squibb Company; CereSpir, Inc.; Cogstate; Eisai Inc.; Elan Pharmaceuticals, Inc.; Eli Lilly and Company; EuroImmun; F. Hoffmann-La Roche Ltd and its affiliated company Genentech, Inc.; Fujirebio; GE Healthcare; IXICO Ltd.; Janssen Alzheimer Immunotherapy Research & Development, LLC.; Johnson & Johnson Pharmaceutical Research & Development LLC.; Lumosity; Lundbeck; Merck & Co., Inc.; Meso Scale Diagnostics, LLC.; NeuroRx Research; Neurotrack Technologies; Novartis Pharmaceuticals Corporation; Pfizer Inc.; Piramal Imaging; Servier; Takeda Pharmaceutical Company; and Transition Therapeutics. The Canadian Institutes of Health Research is providing funds to support ADNI clinical sites in Canada. Private sector contributions are facilitated by the Foundation for the National Institutes of Health (www.fnih.org). The grantee organization is the Northern California Institute for Research and Education, and the study is coordinated by the Alzheimer?s Therapeutic Research Institute at the University of Southern California. ADNI data are disseminated by the Laboratory for Neuro Imaging at the University of Southern California.

## References

Attias, H. (2000). A variational Bayesian framework for graphical models. *Advances in neural information processing systems* **12** 209–215.

Beckmann, C. F. and Smith, S. M. (2004). Probabilistic independent component analysis for functional magnetic resonance imaging. *Medical Imaging, IEEE Transactions on* **23** 137–152.

Beckmann, C. F. and Smith, S. M. (2005). Tensorial extensions of independent component analysis for multisubject FMRI analysis. *Neuroimage* **25** 294–311.

Beckmann, C. F., DeLuca, M., Devlin, J. T. and Smith, S. M. (2005). Investigations into resting-state connectivity using independent component analysis. *Philosophical Transactions of the Royal Society of London B: Biological Sciences* **360** 1001–1013.

Biswal, B. B. and Ulmer, J. L. (1999). Blind source separation of multiple signal sources of fMRI data sets using independent component analysis. *Journal of computer assisted tomography* **23** 265–271.

Bullmore, E., Brammer, M., Williams, S. C., Rabe-Hesketh, S., Janot, N., David, A., Mellers, J., Howard, R. and Sham, P. (1996). Statistical methods of estimation and inference for functional MR image analysis. *Magnetic Resonance in Medicine* **35** 261–277.

Calhoun, V., Adali, T., Pearlson, G. and Pekar, J. (2001a). A method for making group inferences from functional MRI data using independent component analysis. *Human brain mapping* **14** 140–151.

Calhoun, V. D., Adali, T., McGinty, V., Pekar, J. J., Watson, T. and Pearlson, G. (2001b). fMRI activation in a visual-perception task: network of areas detected using the general linear model and independent components analysis. *NeuroImage* **14** 1080–1088.

Chumbley, J. R. and Friston, K. J. (2009). False discovery rate revisited: FDR and topological inference using Gaussian random fields. *Neuroimage* **44** 62–70.

Dai, T., Guo, Y., Initiative, A. D. N. et al. (2017). Predicting individual brain functional connectivity using a Bayesian hierarchical model. *NeuroImage* **147** 772–787.

Daubechies, I., Roussos, E., Takerkart, S., Benharrosh, M., Golden, C., D'ardenne, K., Richter, W., Cohen, J. and Haxby, J. (2009). Independent component analysis for brain fMRI does not select for independence. *Proceedings of the National Academy of Sciences* **106** 10415–10422.

Dempster, A. P., Laird, N. M. and Rubin, D. B. (1977). Maximum likelihood from incomplete data via the EM algorithm. *Journal of the royal statistical society. Series B (methodological)* 1–38.

Dettwiler, A., Murugavel, M., Putukian, M., Cubon, V., Furtado, J. and Osherson, D. (2014). Persistent differences in patterns of brain activation after sports-related concussion: a longitudinal functional magnetic resonance imaging study. *Journal of neurotrauma* **31** 180–188.

Genovese, C. R., Lazar, N. A. and Nichols, T. (2002). Thresholding of statistical maps in functional neuroimaging using the false discovery rate. *Neuroimage* **15** 870–878.

Greicius, M. D., Srivastava, G., Reiss, A. L. and Menon, V. (2004). Default-mode network activity distinguishes Alzheimer's disease from healthy aging: evidence from functional MRI. *Proceedings of the National Academy of Sciences of the United States of America* **101** 4637–4642.

Guo, Y. (2011). A general probabilistic model for group independent component analysis and its estimation methods. *Biometrics* **67** 1532–1542.

Guo, Y. and Pagnoni, G. (2008). A unified framework for group independent component analysis for multi-subject fMRI data. *NeuroImage* **42** 1078–1093.

Guo, Y. and Tang, L. (2013). A hierarchical model for probabilistic independent component analysis of multi-subject fMRI studies. *Biometrics* **69** 970–981.

Himberg, J., Hyvärinen, A. and Esposito, F. (2004). Validating the independent components of neuroimaging time series via clustering and visualization. *Neuroimage* **22** 1214–1222.

Hyvärinen, A., Karhunen, J. and Oja, E. (2001). *Independent component analysis* **46**. John Wiley & Sons.

Hyvärinen, A. and Oja, E. (2000). Independent component analysis: algorithms and applications. *Neural networks* **13** 411–430.

Kemmer, P. B., Guo, Y., Wang, Y. and Pagnoni, G. (2015). Network-based characterization of brain functional connectivity in Zen practitioners. *Frontiers in psychology* **6**.

Kostantinos, N. (2000). Gaussian mixtures and their applications to signal processing. *Advanced Signal Processing Handbook: Theory and Implementation for Radar, Sonar, and Medical Imaging Real Time Systems.*

Lee, S., Zipunnikov, V., Reich, D. S. and Pham, D. L. (2015). Statistical image analysis of longitudinal RAVENS images. *Frontiers in neuroscience* **9** 368.

Li, Y., Zhu, H., Chen, Y., An, H., Gilmore, J., Lin, W. and Shen, D. (2009). LSTGEE: Longitudinal analysis of neuroimaging data. In *Medical Imaging 2009: Image Processing* **7259** 72590F. International Society for Optics and Photonics.

Lukemire, J., Wang, Y., Verma, A. and Guo, Y. (2018). HINT: A Toolbox for Hierarchical Modeling of Neuroimaging Data. *arXiv preprint arXiv:1803.07587.*

Mckeown, M. J., Makeig, S., Brown, G. G., Jung, T.-P., Kindermann, S. S., Kindermann, R. S., Bell, A. J. and Sejnowski, T. J. (1998). Analysis of fMRI Data by Blind Separation Into Independent Spatial Components. *Human Brain Mapping* **6** 160–188.

McLachlan, G. and Peel, D. (2004). *Finite mixture models.* John Wiley & Sons.

Minka, T. P. (2000). Automatic choice of dimensionality for PCA. In *NIPS* **13** 598–604.

Satterthwaite, T. D., Wolf, D. H., Roalf, D. R., Ruparel, K., Erus, G., Vandekar, S., Gennatas, E. D., Elliott, M. A., Smith, A., Hakonarson, H. et al. (2014). Linked sex differences in cognition and functional connectivity in youth. *Cerebral cortex* **25** 2383–2394.

Shi, R. and Guo, Y. (2016). Investigating differences in brain functional networks using hierarchical covariate-adjusted independent component analysis. *The annals of applied statistics* **10** 1930.

Smith, S. M., Fox, P. T., Miller, K. L., Glahn, D. C., Fox, P. M., Mackay, C. E., Filippini, N., Watkins, K. E., Toro, R., Laird, A. R. et al. (2009). Correspondence of the brain's functional architecture during activation and rest. *Proceedings of the National Academy of Sciences* **106** 13040–13045.

Storey, J. D. (2011). False discovery rate. In *International encyclopedia of statistical science* 504–508. Springer.

Wang, Y., Wu, H. and Yu, T. (2017). Differential gene network analysis from single cell RNA-seq. *Journal of Genetics and Genomics* **44** 331–334.

Wang, Y., Zhao, Y., Zhang, L., Liang, J., Zeng, M. and Liu, X. (2013). Graph construction based on re-weighted sparse representation for semi-supervised learning. *Journal of Information & Computational Science* **10** 375–383.

Wang, Y., Kang, J., Kemmer, P. B. and Guo, Y. (2016). An efficient and reliable statistical method for estimating functional connectivity in large scale brain networks using partial correlation. *Frontiers in neuroscience* **10**.

Wu, K., Taki, Y., Sato, K., Qi, H., Kawashima, R. and Fukuda, H. (2013). A longitudinal

study of structural brain network changes with normal aging. *Frontiers in human neuroscience* **7** 113.

Xu, L., Cheung, C., Yang, H. and Amari, S. (1997). Maximum equalization by entropy maximization and mixture of cumulative distribution functions. In *Proc. of ICNN97* 1821–1826.

Zhao, X.-H., Wang, P.-J., Li, C.-B., Hu, Z.-H., Xi, Q., Wu, W.-Y. and Tang, X.-W. (2007). Altered default mode network activity in patient with anxiety disorders: an fMRI study. *European Journal of Radiology* **63** 373–378.

## Appendix

*1. Q-function in E step:* The detailed expression for the complete data log-likelihood function at each voxel $v$ is:

$$l_v(\Theta) = \sum_{i=1}^{N}\sum_{j=1}^{K}\Big[\log g\left(\boldsymbol{y}_{ij}(v); \boldsymbol{A}_{ij}\boldsymbol{s}_{ij}(v), \boldsymbol{E}\right) + \log g\left(\boldsymbol{s}_{ij}(v); \boldsymbol{s}_0(v) + \boldsymbol{b}_i(v) + \boldsymbol{C}_j(v)\boldsymbol{x}_i^*, \tau^2\boldsymbol{I}\right)\Big] + \sum_{i=1}^{N}\log g\left(\boldsymbol{b}_i(v); \boldsymbol{0}, \boldsymbol{D}\right) + \log g\left(\boldsymbol{s}_0(v); \boldsymbol{\mu}_{\boldsymbol{z}(v)}, \boldsymbol{\Sigma}_{\boldsymbol{z}(v)}\right) + \sum_{\ell=1}^{q}\log \pi_{l,\boldsymbol{z}_l(v)}$$

where $\boldsymbol{C}_j(v) = [\boldsymbol{\alpha}_j(v), \boldsymbol{\beta}_j(v)']$ of dimension $q \times (p+1)$, $\boldsymbol{x}_i^* = [1, \boldsymbol{x}_i']'$ and $g(\boldsymbol{x}; \boldsymbol{\mu}, \boldsymbol{\Sigma})$ denotes the pdf of multivariate normal distribution for random vector $x$ with mean $\boldsymbol{\mu}$ and covariance $\boldsymbol{\Sigma}$.

We derive the Q function in E step as follows,

$$\begin{aligned} Q(\Theta|\hat{\Theta}^{(k)}) &= E[l(\Theta; \mathcal{Y}, \mathcal{X}, \mathcal{S}, \mathcal{B}, \mathcal{Z})|\mathcal{Y}] \\ &= Q_1(\Theta|\hat{\Theta}^{(k)}) + Q_2(\Theta|\hat{\Theta}^{(k)}) + Q_3(\Theta|\hat{\Theta}^{(k)}) + Q_4(\Theta|\hat{\Theta}^{(k)}) + Q_5(\Theta|\hat{\Theta}^{(k)}), \end{aligned}$$

where

$$Q_1(\Theta|\hat{\Theta}^{(k)}) = -\frac{NKV}{2}\log|\boldsymbol{E}| - \frac{1}{2}\sum_{v=1}^{V}\sum_{i=1}^{N}\sum_{j=1}^{K}\mathrm{tr}\Big\{\Big[\boldsymbol{y}_{ij}(v)\boldsymbol{y}_{ij}(v)' - 2\boldsymbol{A}_{ij}E[\boldsymbol{s}_{ij}(v)|\mathbf{y}(v); \hat{\Theta}^{(k)}]\boldsymbol{y}_{ij}(v)' + \boldsymbol{A}_{ij}E[\boldsymbol{s}_{ij}(v)\boldsymbol{s}_{ij}(v)'|\mathbf{y}(v); \hat{\Theta}^{(k)}]\boldsymbol{A}_{ij}'\Big]\boldsymbol{E}^{-1}\Big\},$$

$$Q_2(\Theta|\hat{\Theta}^{(k)}) = -\frac{NKVq}{2}\log|\tau^2| - \frac{1}{2\tau^2}\sum_{v=1}^{V}\sum_{i=1}^{N}\sum_{j=1}^{K}\mathrm{tr}\Big\{\Big[E[\boldsymbol{s}_{ij}(v)\boldsymbol{s}_{ij}(v)' + \boldsymbol{s}_0(v)\boldsymbol{s}_0(v)' + \boldsymbol{b}_i(v)\boldsymbol{b}_i(v)'|\mathbf{y}(v); \hat{\Theta}^{(k)}] + 2E[\boldsymbol{b}_i(v)\boldsymbol{s}_0(v)'|\mathbf{y}(v); \hat{\Theta}^{(k)}] + 2\boldsymbol{x}_i^{*\prime}\boldsymbol{C}_j(v)'E[\boldsymbol{s}_0(v) + \boldsymbol{b}_i(v) - \boldsymbol{s}_{ij}(v)|\mathbf{y}(v); \hat{\Theta}^{(k)}] + \boldsymbol{C}_j(v)\boldsymbol{x}_i^*\boldsymbol{x}_i^{*\prime}\boldsymbol{C}_j(v)' - 2E[\boldsymbol{s}_0(v)\boldsymbol{s}_{ij}(v)' + \boldsymbol{b}_i(v)\boldsymbol{s}_{ij}(v)'|\mathbf{y}(v); \hat{\Theta}^{(k)}]\Big]\Big\},$$

$$Q_3(\Theta|\hat{\Theta}^{(k)}) = -\frac{NV}{2}\log|\boldsymbol{D}| - \frac{1}{2}\sum_{v=1}^{V}\sum_{i=1}^{N}\mathrm{tr}\Big\{\boldsymbol{D}^{-1}E[\boldsymbol{b}_i(v)\boldsymbol{b}_i(v)'|\mathbf{y}(v); \hat{\Theta}^{(k)}]\Big\},$$

$$Q_4(\Theta|\hat{\Theta}^{(k)}) = -\frac{1}{2}\sum_{v=1}^{V}\sum_{\ell=1}^{q}\sum_{j=1}^{m} p[z_\ell(v)=j|\mathbf{y}(v);\hat{\Theta}^{(k)}]\Big\{\log\sigma_{\ell,j}^2 + \frac{1}{\sigma_{\ell,j}^2}\Big[\mu_{\ell,j}^2 + E[s_0^{(\ell)}(v)^2|z_\ell(v)=j;\mathbf{y}(v),\hat{\Theta}^{(k)}] - 2\mu_{\ell,j}E[s_0^{(\ell)}(v)|z_\ell(v)=j,\mathbf{y}(v);\hat{\Theta}^{(k)}]\Big]\Big\},$$

$$Q_5(\Theta|\hat{\Theta}^{(k)}) = \sum_{v=1}^{V}\sum_{\ell=1}^{q}\sum_{j=1}^{m} p[z_\ell(v)=j|\mathbf{y}(v);\hat{\Theta}^{(k)}]\log\pi_{\ell,j},$$

*2. Details about the E step of the exact EM algorithm.* In this section, we provide the details about the derivation in the exact E step. By collapsing our model across N subjects and K visits, for $v = 1, .., V$,

$$\begin{aligned}(11)\qquad \boldsymbol{y}(v) &= \boldsymbol{A}\left(\boldsymbol{U}^{(c)}\boldsymbol{\mu}_{\boldsymbol{z}(v)} + \boldsymbol{U}^{(c)}\boldsymbol{\psi}(v) + \boldsymbol{H}\boldsymbol{b}(v) + \boldsymbol{C}^*(v)\boldsymbol{X}^* + \gamma(v)\right) + \boldsymbol{e}(v),\\ &= \boldsymbol{A}\boldsymbol{U}^{(c)}\boldsymbol{\mu}_{\boldsymbol{z}(v)} + \boldsymbol{A}\boldsymbol{C}^*(v)\boldsymbol{X}^* + \boldsymbol{A}\boldsymbol{R}\boldsymbol{r}_{\boldsymbol{z}(v)} + \boldsymbol{e}(v),\end{aligned}$$

where $\boldsymbol{A} = blockdiag(\boldsymbol{A}_{11}, .., \boldsymbol{A}_{NK})$, $\boldsymbol{b}(v) = [\boldsymbol{b}_1(v)', .., b_N(v)']'$, $\boldsymbol{\gamma}(v) = [\boldsymbol{\gamma}_{11}(v)', .., \boldsymbol{\gamma}_{NK}(v)']'$, $\boldsymbol{e}(v) = [\boldsymbol{e}_{11}(v)', .., \boldsymbol{e}_{NK}(v)']'$, $\boldsymbol{U}^{(c)} = \mathbf{1}_{NK}\otimes\boldsymbol{I}_q$, $\boldsymbol{H} = (\boldsymbol{I}_N\otimes\mathbf{1}_K)\otimes\boldsymbol{I}_q$, $\boldsymbol{C}^*(v) = \boldsymbol{I}_N\otimes[\boldsymbol{C}_1(v)', .., \boldsymbol{C}_K(v)']'$, $\boldsymbol{X}^* = [\boldsymbol{x}_1^{*\prime}, ..., \boldsymbol{x}_N^{*}{}']'$ $\boldsymbol{R} = [\boldsymbol{H}, \boldsymbol{U}^{(c)}, \boldsymbol{I}_{qNK}]$, $\boldsymbol{r}_{\boldsymbol{z}(v)} = [\boldsymbol{b}(v)', \boldsymbol{\psi}'_{\boldsymbol{z}(v)}, \gamma(v)']'$. Conditioned on latent variable $\boldsymbol{z}(v)$, (11) can be represented as:

$$\begin{aligned}(12)\qquad \boldsymbol{y}(v) - \boldsymbol{A}\boldsymbol{U}^{(c)}\boldsymbol{\mu}_{\boldsymbol{z}(v)} - \boldsymbol{A}\boldsymbol{C}^*(v)\boldsymbol{X}^*\Big|\boldsymbol{r}_{\boldsymbol{z}(v)}, \boldsymbol{z}(v) &\sim N(\boldsymbol{A}\boldsymbol{R}\boldsymbol{r}_{\boldsymbol{z}(v)}, \boldsymbol{\Upsilon}_v),\\ \boldsymbol{r}_{\boldsymbol{z}(v)}|\boldsymbol{z}(v) &\sim N(\mathbf{0}, \boldsymbol{\Gamma}_{\boldsymbol{z}(v)})\end{aligned}$$

where $\boldsymbol{\Upsilon}_v = \boldsymbol{I}_{NK}\otimes\boldsymbol{E}_v$, $\boldsymbol{\Gamma}_{\boldsymbol{z}(v)} = blockdiag(\boldsymbol{I}_N\otimes\boldsymbol{D}, \boldsymbol{\Sigma}_{\boldsymbol{z}(v)}, \tau^2\boldsymbol{I}_{qNK})$. From (12), we can derive the conditional distribution of $[\boldsymbol{r}_{\boldsymbol{z}(v)}|\boldsymbol{y}(v), \boldsymbol{z}(v)]$ through Bayes' Theorem,

$$\begin{aligned}\boldsymbol{r}_{\boldsymbol{z}(v)}\Big|\boldsymbol{y}(v), \boldsymbol{z}(v) &\sim N(\boldsymbol{\mu}_{\boldsymbol{r}(v)|\boldsymbol{y}(v)}, \boldsymbol{\Sigma}_{\boldsymbol{r}(v)|\boldsymbol{y}(v)}),\\ \boldsymbol{\mu}_{\boldsymbol{r}(v)|\boldsymbol{y}(v)} &= \boldsymbol{\Sigma}_{\boldsymbol{r}(v)|\boldsymbol{y}(v)}\boldsymbol{R}'\boldsymbol{A}'\boldsymbol{\Upsilon}^{-1}\Big[\boldsymbol{y}(v) - \boldsymbol{A}\boldsymbol{U}^{(c)}\boldsymbol{\mu}_{\boldsymbol{z}(v)} - \boldsymbol{A}\boldsymbol{C}^*(v)\boldsymbol{X}^*\Big],\\ \boldsymbol{\Sigma}_{\boldsymbol{r}(v)|\boldsymbol{y}(v)} &= \left(\boldsymbol{\Gamma}_{\boldsymbol{z}(v)}^{-1} + \boldsymbol{R}'\boldsymbol{A}'\boldsymbol{\Upsilon}^{-1}\boldsymbol{A}\boldsymbol{R}\right)^{-1}.\end{aligned}$$

Next, we evaluate the conditional distribution of $\boldsymbol{L}(v)$. Given that $\boldsymbol{L}(v) = \boldsymbol{P}\boldsymbol{r}_{\boldsymbol{z}(v)} + \boldsymbol{Q}_{\boldsymbol{z}(v)}$, we have $\boldsymbol{L}(v)\Big|\boldsymbol{y}(v), \boldsymbol{z}(v) \sim N(\boldsymbol{P}\boldsymbol{\mu}_{\boldsymbol{r}(v)|\boldsymbol{y}(v)} + \boldsymbol{Q}_{\boldsymbol{z}(v)}, \boldsymbol{P}\boldsymbol{\Sigma}_{\boldsymbol{r}(v)|\boldsymbol{y}(v)}\boldsymbol{P}')$, where

$$\boldsymbol{P} = \begin{pmatrix}\boldsymbol{I}_{qN} & \mathbf{0} & \mathbf{0}\\ \mathbf{0} & \boldsymbol{I}_q & \mathbf{0}\\ \boldsymbol{H} & \boldsymbol{U}^{(c)} & \boldsymbol{I}_{qNK}\end{pmatrix}, \boldsymbol{Q}_{\boldsymbol{z}(v)} = \begin{pmatrix}\mathbf{0}\\ \mu_{\boldsymbol{z}(v)}\\ \boldsymbol{U}^{(c)}\mu_{\boldsymbol{z}(v)} + \boldsymbol{C}^*(v)\boldsymbol{X}^*\end{pmatrix}.$$

Based on Bayes' Theorem, we have

$$p[\boldsymbol{z}(v)\mid\boldsymbol{y}(v)] \propto \Big(\prod_{l=1}^{q}\pi_{l,z_l(v)}\Big)g(\boldsymbol{A}\boldsymbol{U}^{(c)}\boldsymbol{\mu}_{\boldsymbol{z}(v)} + \boldsymbol{A}\boldsymbol{C}^*(v)\boldsymbol{X}^*, \boldsymbol{A}\boldsymbol{R}\boldsymbol{\Gamma}_{\boldsymbol{z}(v)}\boldsymbol{R}'\boldsymbol{A}' + \boldsymbol{\Upsilon}_v).$$

By integrating out $p[\boldsymbol{z}(v)\mid\boldsymbol{y}(v)]$, we obtain the conditional distribution of $\boldsymbol{L}(v)$.

*3. Details about the M step of the exact EM* In this section, we only provide the M step of the exact EM.

- Update the time-specific covariate effects $\boldsymbol{C}_j(v)$: for $j = 1, .., K$, $v = 1, .., V$,

$$\hat{\boldsymbol{C}}_j(v)^{(k+1)} = \left(\sum_{i=1}^{N} \boldsymbol{x}_i^* \boldsymbol{x}_i^{*\prime}\right)^{-1} \sum_{i=1}^{N} \left\{ \boldsymbol{x}_i^* \left( E[\boldsymbol{s}_{ij}(v)' - \boldsymbol{s}_0(v)' - \boldsymbol{b}_0(v)' | \mathbf{y}(v); \hat{\Theta}^{(k)}] \right) \right\}.$$

- Update the mixing matrices $\boldsymbol{A}_{ij}$: for $i = 1, .., N$, $j = 1, .., K$,

$$\check{\boldsymbol{A}}_{ij}^{(k+1)} = \left\{ \sum_{v=1}^{V} \boldsymbol{y}_{ij}(v) E[\boldsymbol{s}_{ij}(v)' | \mathbf{y}(v); \hat{\Theta}^{(k)}] \right\} \left\{ \sum_{v=1}^{V} E[\boldsymbol{s}_{ij}(v) \boldsymbol{s}_{ij}(v)' | \mathbf{y}(v); \hat{\Theta}^{(k)}] \right\}^{-1},$$

  and then update $\widehat{\boldsymbol{A}}_{ij}^{(k+1)} = \mathcal{H}(\check{\boldsymbol{A}}_{ij}^{(k+1)})$ where $\mathcal{H}(\cdot)$ is the orthogonalization transformation.
- Update the first level variance term $\boldsymbol{E}_v = \sigma_0^2 \boldsymbol{I}_q$ with:

$$\begin{aligned}\hat{\sigma}_0^{2(k+1)} =& \frac{1}{NKVq} \sum_{v=1}^{V} \sum_{i=1}^{N} \sum_{j=1}^{K} \Big\{ \boldsymbol{y}_{ij}(v)' \boldsymbol{y}_{ij}(v) - 2 \boldsymbol{y}_{ij}(v)' \widehat{\boldsymbol{A}}_{ij}^{(k+1)} E[\boldsymbol{s}_{ij}(v) | \mathbf{y}(v); \hat{\Theta}^{(k)}] \\ & + \mathrm{tr} \left[ \widehat{\boldsymbol{A}}_{ij}^{(k+1)} E[\boldsymbol{s}_{ij}(v) \boldsymbol{s}_{ij}(v)' | \mathbf{y}(v); \hat{\Theta}^{(k)}] \widehat{\boldsymbol{A}}_{ij}^{(k+1)\prime} \right] \Big\}.\end{aligned}$$

- Update subject-specific variance term $\boldsymbol{D}$ :

$$\hat{\boldsymbol{D}}^{(k+1)} = \frac{1}{NV} \sum_{v=1}^{V} \sum_{i=1}^{N} E[\boldsymbol{b}_i(v) \boldsymbol{b}_i(v)' | \mathbf{y}(v); \hat{\Theta}^{(k)}],$$

- Update second level variance term $\tau^2 \boldsymbol{I}_q$ :

$$\begin{aligned}\hat{\tau}^{2(k+1)} =& \frac{1}{NKVq} \sum_{v=1}^{V} \sum_{i=1}^{N} \sum_{j=1}^{K} \mathrm{tr} \Big\{ E[\boldsymbol{s}_{ij}(v) \boldsymbol{s}_{ij}(v)' + \boldsymbol{s}_0(v) \boldsymbol{s}_0(v)' + \boldsymbol{b}_i(v) \boldsymbol{b}_i(v)' | \mathbf{y}(v); \hat{\Theta}^{(k)}] \\ & + 2E[\boldsymbol{b}_i(v) \boldsymbol{s}_0(v)' | \mathbf{y}(v); \hat{\Theta}^{(k)}] + 2 \boldsymbol{x}_i^{*\prime} \boldsymbol{C}_j(v)' E[\boldsymbol{s}_0(v) + \boldsymbol{b}_i(v) - \boldsymbol{s}_{ij}(v) | \mathbf{y}(v); \hat{\Theta}^{(k)}] \\ & + \boldsymbol{C}_j(v) \boldsymbol{x}_i^* \boldsymbol{x}_i^{*\prime} \boldsymbol{C}_j(v)' - 2E[\boldsymbol{s}_0(v) \boldsymbol{s}_{ij}(v)' + \boldsymbol{b}_i(v) \boldsymbol{s}_{ij}(v)' | \mathbf{y}(v); \hat{\Theta}^{(k)}] \Big\},\end{aligned}$$

- Update $\pi_{\ell,j}$:

$$\hat{\pi}_{\ell,j}^{(k+1)} = \frac{1}{V} \sum_{v=1}^{V} p[\boldsymbol{z}_\ell(v) = j | \mathbf{y}(v); \hat{\Theta}^{(k)}].$$

- Update $\mu_{\ell,j}$:

$$\hat{\mu}_{\ell,j}^{(k+1)} = \frac{\sum_{v=1}^{V} p[\boldsymbol{z}_\ell(v) = j | \mathbf{y}(v); \hat{\Theta}^{(k)}] E[s_{0\ell}(v) | \boldsymbol{z}_\ell(v) = j, \mathbf{y}(v); \hat{\Theta}^{(k)}]}{V \hat{\pi}_{\ell,j}^{(k+1)}}.$$

- Update $\sigma^2_{\ell,j}$ :

$$\hat{\sigma}^{2(k+1)}_{\ell,j} = \frac{\sum_{v=1}^{V} p[\boldsymbol{z}_\ell(v)=j|\mathbf{y}(v);\hat{\Theta}^{(k)}]E[s_{0\ell}(v)^2|\boldsymbol{z}_\ell(v)=j,\mathbf{y}(v);\hat{\Theta}^{(k)}]}{V\hat{\pi}^{(k+1)}_{\ell,j}} - [\hat{\mu}^{(k+1)}_{\ell,j}]^2.$$

Here, $E[s_{0\ell}(v) \mid z_\ell(v)=j,\mathbf{y}(v);\Theta]$, $E[s_{0\ell}(v)^2 \mid z_\ell(v)=j,\mathbf{y}(v);\Theta]$ and $p[z_\ell(v)=j \mid \mathbf{y}(v);\Theta]$ are the marginal conditional moments and probability related to the $\ell$th IC. They are derived by summing across all the possible states of the other $q-1$ ICs as follows,

(13)
$$E[s_{0\ell}(v) \mid z_\ell(v)=j,\mathbf{y}(v);\Theta] = \frac{\sum_{\boldsymbol{z}(v)\in\mathcal{R}^{(\ell,j)}} p[\boldsymbol{z}(v) \mid \mathbf{y}(v);\Theta]E[s_{0\ell}(v) \mid \mathbf{y}(v),\boldsymbol{z}(v);\Theta]}{p[z_\ell(v)=j \mid \mathbf{y}(v);\Theta]},$$

$$p[z_\ell(v)=j \mid \mathbf{y}(v);\Theta] = \sum_{\boldsymbol{z}(v)\in\mathcal{R}^{(\ell,j)}} p[\boldsymbol{z}(v) \mid \mathbf{y}(v);\Theta].$$

where $\mathcal{R}^{(\ell,j)}$ is defined as $\{\boldsymbol{z}^r \in \mathcal{R} : z^r_\ell = j\}$ for all $\ell = 1,..,q, j = 1,...,m$.

*4. Statistical inference for testing covariate effects in L-ICA:* In this section, we present the statistical inference procedure for testing covariate effects in L-ICA. We first stack the fMRI data from all visits of a subject to have the subject-specific fMRI data $\boldsymbol{y}_i(v)$ of dimension $qK\times 1$ which is $[\boldsymbol{y}_{i1}(v)',...,\boldsymbol{y}_{iK}(v)']'$, and a non-hierarchical form of L-ICA is derived by combining equations (1),(2) and (3),

$$\boldsymbol{A}_i'\boldsymbol{y}_i(v) = \boldsymbol{U}\boldsymbol{\mu}_{\boldsymbol{z}(v)} + \boldsymbol{\alpha}(v) + \boldsymbol{X}_i\boldsymbol{\beta}(v) + \boldsymbol{U}\boldsymbol{\psi}_{\boldsymbol{z}(v)} + \boldsymbol{U}\boldsymbol{b}_i(v) + \boldsymbol{\gamma}_i(v) + \boldsymbol{A}_i'\boldsymbol{e}_i(v), \tag{14}$$

where $\boldsymbol{A}_i = blkdiag(\boldsymbol{A}_{i1},...,\boldsymbol{A}_{iK})$, $\boldsymbol{\gamma}_i(v) = [\boldsymbol{\gamma}_{i1}(v)',...,\boldsymbol{\gamma}_{iK}(v)']'$, $\boldsymbol{e}_i(v) = [\boldsymbol{e}_{i1}(v)',...,\boldsymbol{e}_{iK}(v)']'$, $\boldsymbol{\alpha}(v) = [\boldsymbol{\alpha}_1(v)',\boldsymbol{\alpha}_2(v)',..,\boldsymbol{\alpha}_K(v)']'$, $\boldsymbol{\beta}(v) = [\text{vec}\,[\boldsymbol{\beta}_1(v)']',...,\text{vec}\,[\boldsymbol{\beta}_K(v)']']'$, $\boldsymbol{U} = \mathbf{1}_K \otimes \boldsymbol{I}_q$ and $\boldsymbol{X}_i = \boldsymbol{I}_K \otimes (\boldsymbol{x}_i' \otimes \boldsymbol{I}_q)$. The model in (14) is further re-written as

$$\begin{aligned}\boldsymbol{y}^*_i(v) &= \boldsymbol{X}_0\boldsymbol{\alpha}^*(v) + \boldsymbol{X}_i\boldsymbol{\beta}(v) + \boldsymbol{\zeta}_i(v),\\ &= \boldsymbol{X}^*_i\boldsymbol{C}^*(v) + \boldsymbol{\zeta}_i(v),\end{aligned} \tag{15}$$

where $\boldsymbol{y}^*_i(v) = \boldsymbol{A}_i'\boldsymbol{y}_i(v)$, $\boldsymbol{X}^*_i = [\boldsymbol{X}_0,\boldsymbol{X}_i]$, $\boldsymbol{X}_0 = \begin{pmatrix} 1 & \mathbf{0}'_{K-1} \\ \mathbf{1}_{K-1} & \boldsymbol{I}_{K-1}\end{pmatrix} \otimes \boldsymbol{I}_q$, $\boldsymbol{\alpha}^*(v) = [\boldsymbol{\mu}'_{\boldsymbol{z}(v)},\boldsymbol{\alpha}_2(v)',..,\boldsymbol{\alpha}_K(v)']'$, $\boldsymbol{C}^*(v) = [\boldsymbol{\alpha}^*(v)',\boldsymbol{\beta}(v)']'$ and $\boldsymbol{\zeta}_i(v) = \boldsymbol{U}\boldsymbol{\psi}_{\boldsymbol{z}(v)} + \boldsymbol{U}\boldsymbol{b}_i(v) + \boldsymbol{\gamma}_i(v) + \boldsymbol{A}_i'\boldsymbol{e}_i(v) \sim N(\mathbf{0},\boldsymbol{W}_i(v))$ is the multivariate zero-mean Gaussian noise term where $\boldsymbol{W}_i(v) = \boldsymbol{U}(\boldsymbol{\Sigma}_{\boldsymbol{z}(v)} + \boldsymbol{D})\boldsymbol{U}' + \boldsymbol{A}_i\boldsymbol{E}_v\boldsymbol{A}_i' + \tau^2\boldsymbol{I}_{qK}$, which can be shown as $\boldsymbol{W}_i(v) = \boldsymbol{W}(v) = \boldsymbol{U}(\boldsymbol{\Sigma}_{\boldsymbol{z}(v)} + \boldsymbol{D})\boldsymbol{U}' + (\sigma_0^2+\tau^2)\boldsymbol{I}_{qNK}$.

*5. Details about the pre-processing step prior to L-ICA:* In this section, we provide details about the pre-whitening procedure prior to ICA which is the same as our previous work (Shi and Guo, 2016; Guo, 2011; Guo and Pagnoni, 2008; Guo and Tang, 2013). First we assume that each column of $\widetilde{\boldsymbol{Y}}_{ij}$ is centered. The following dimension reduction and whitening procedure is conducted:

$$\boldsymbol{Y}_{ij} = (\boldsymbol{\Lambda}_{ij,q} - \tilde{\sigma}^2_{ij,q}\boldsymbol{I}_q)^{-\frac{1}{2}}\boldsymbol{U}'_{ij,q}\widetilde{\boldsymbol{Y}}_{ij}, \tag{16}$$

where $\boldsymbol{U}_{ij,q}$ and $\boldsymbol{\Lambda}_{ij,q}$ contain the first $q$ eigenvectors and eigenvalues based on the singular value decomposition of $\widetilde{\boldsymbol{Y}}_{ij}$. The residual variance, $\tilde{\sigma}^2_{ij,q}$, is the average of the smallest $T-q$ eigenvalues that are not included in $\boldsymbol{\Lambda}_{ij,q}$ representing the variability in $\widetilde{\boldsymbol{Y}}_{ij}$ that is not accounted by the first $q$ components. The parameter $q$, which is the number of ICs, can be determined using the Laplace approximation method (Minka, 2000).

Department of Biostatistics and Bioinformatics
Rollins School of Public Health
Emory University
1518 Clifton Rd.
Atlanta, Georgia 30322 USA
E-mail: ywan566@emory.edu
yguo2@emory.edu (Corresponding Author)

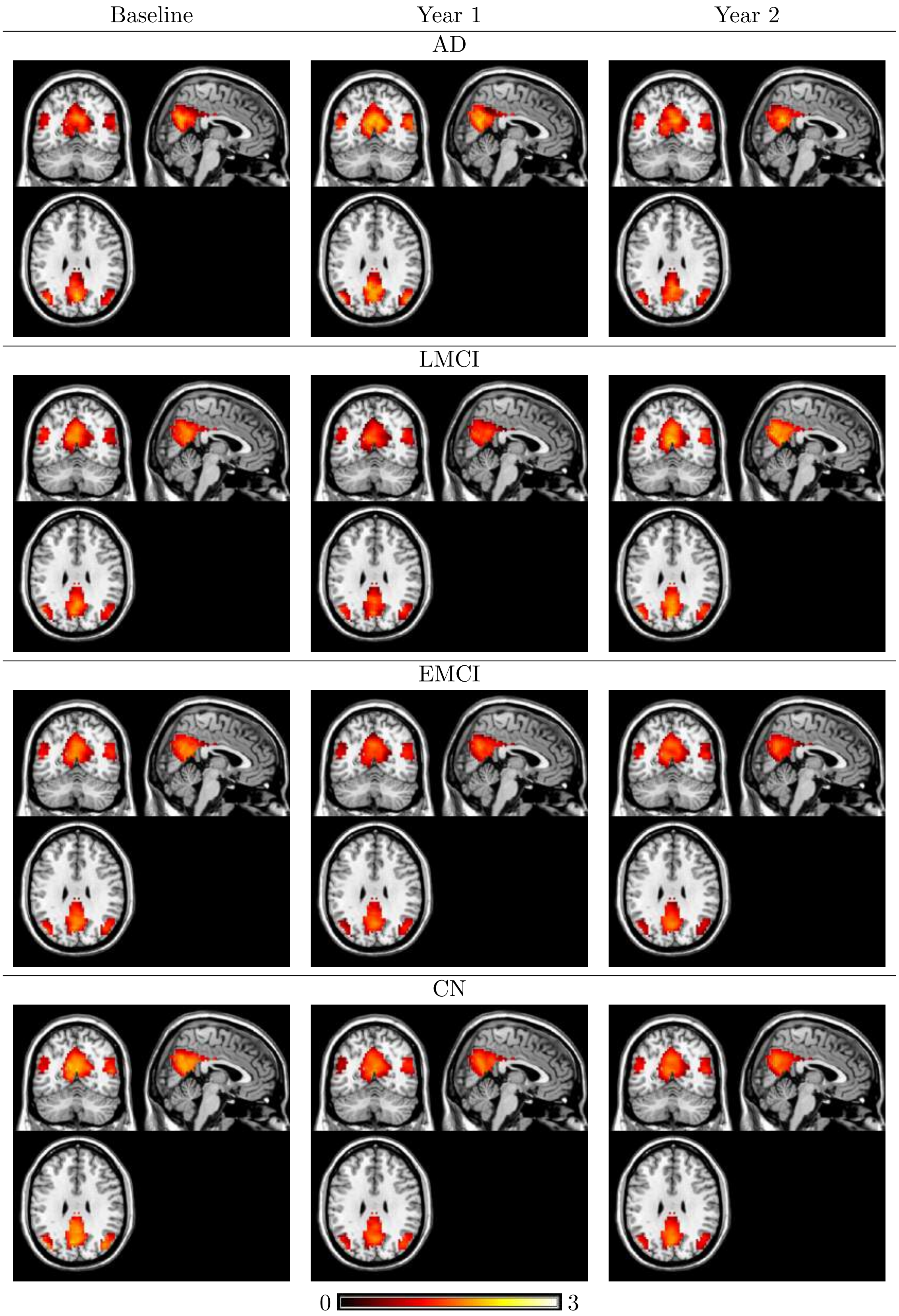

Fig 4. *L-ICA estimates of subpopulation spatial source signal maps for the DMN for the four disease group across the visits, with the mean baseline age (73.7 year old) and are averaged between genders. All IC maps are thresholded based on the source signal intensity level.*

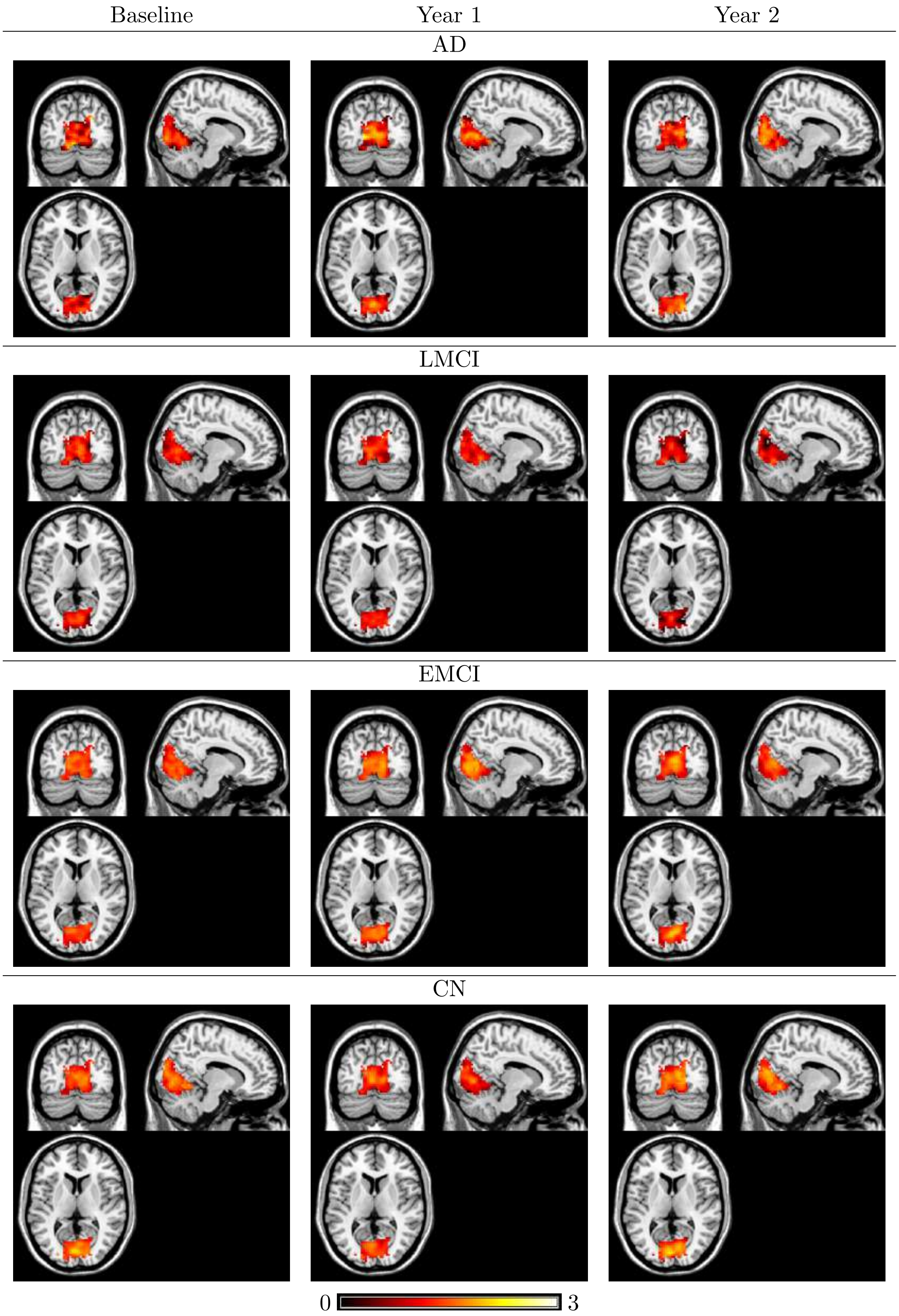

Fig 5. *L-ICA estimates of subpopulation spatial source signal maps for the medial visual network for the four disease group across the visits, with the mean baseline age (73.7 year old) and are averaged between genders. All IC maps are thresholded based on the source signal intensity level.*

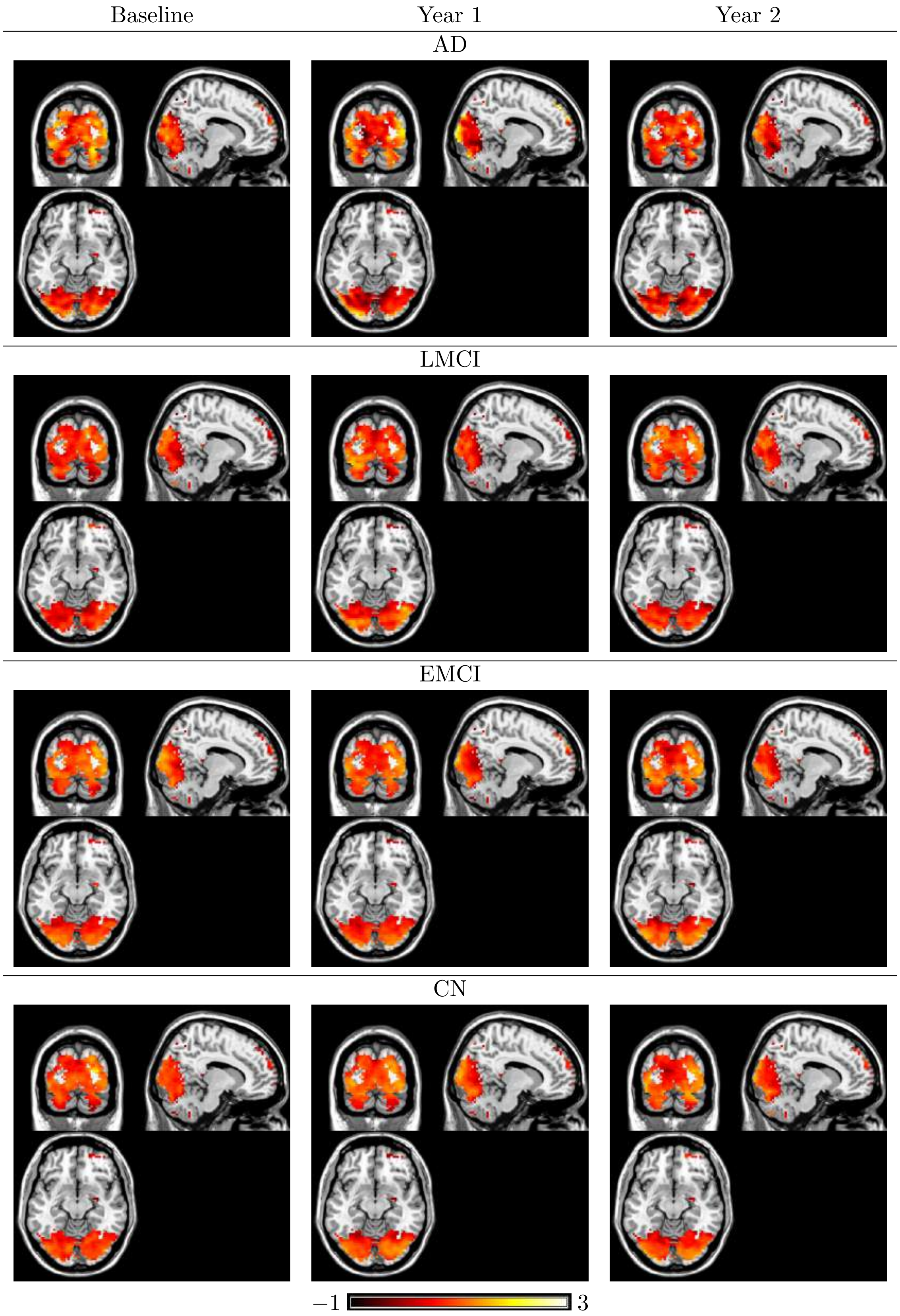

FIG 6. *L-ICA estimates of subpopulation spatial source signal maps for the occipital visual network for the four disease group across the visits, with the mean baseline age (73.7 year old) and are averaged between genders. All IC maps are thresholded based on the source signal intensity level.*

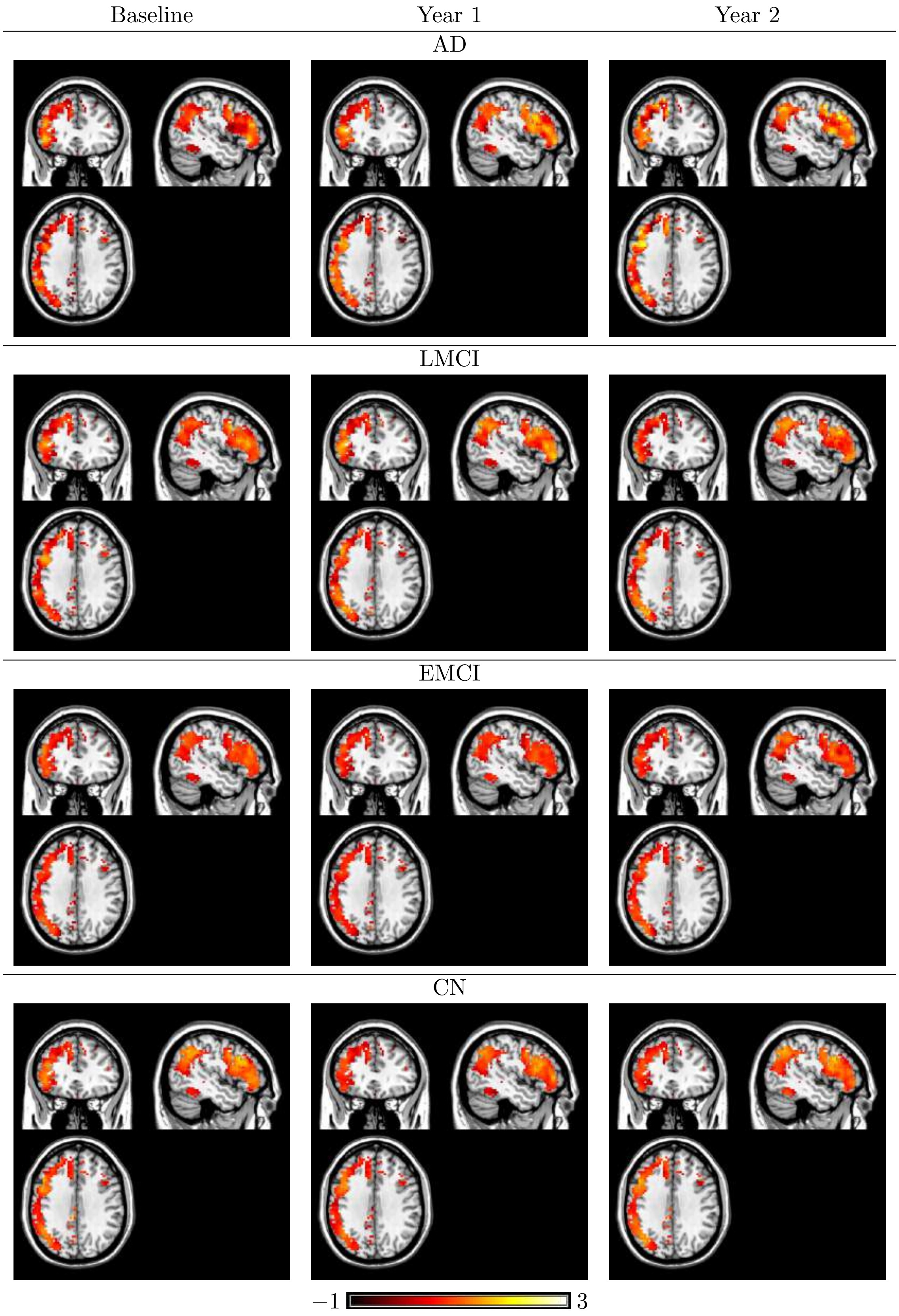

FIG 7. *L-ICA estimates of subpopulation spatial source signal maps for the FPL for the four disease group across the visits, with the mean baseline age (73.7 year old) and are averaged between genders. All IC maps are thresholded based on the source signal intensity level.*

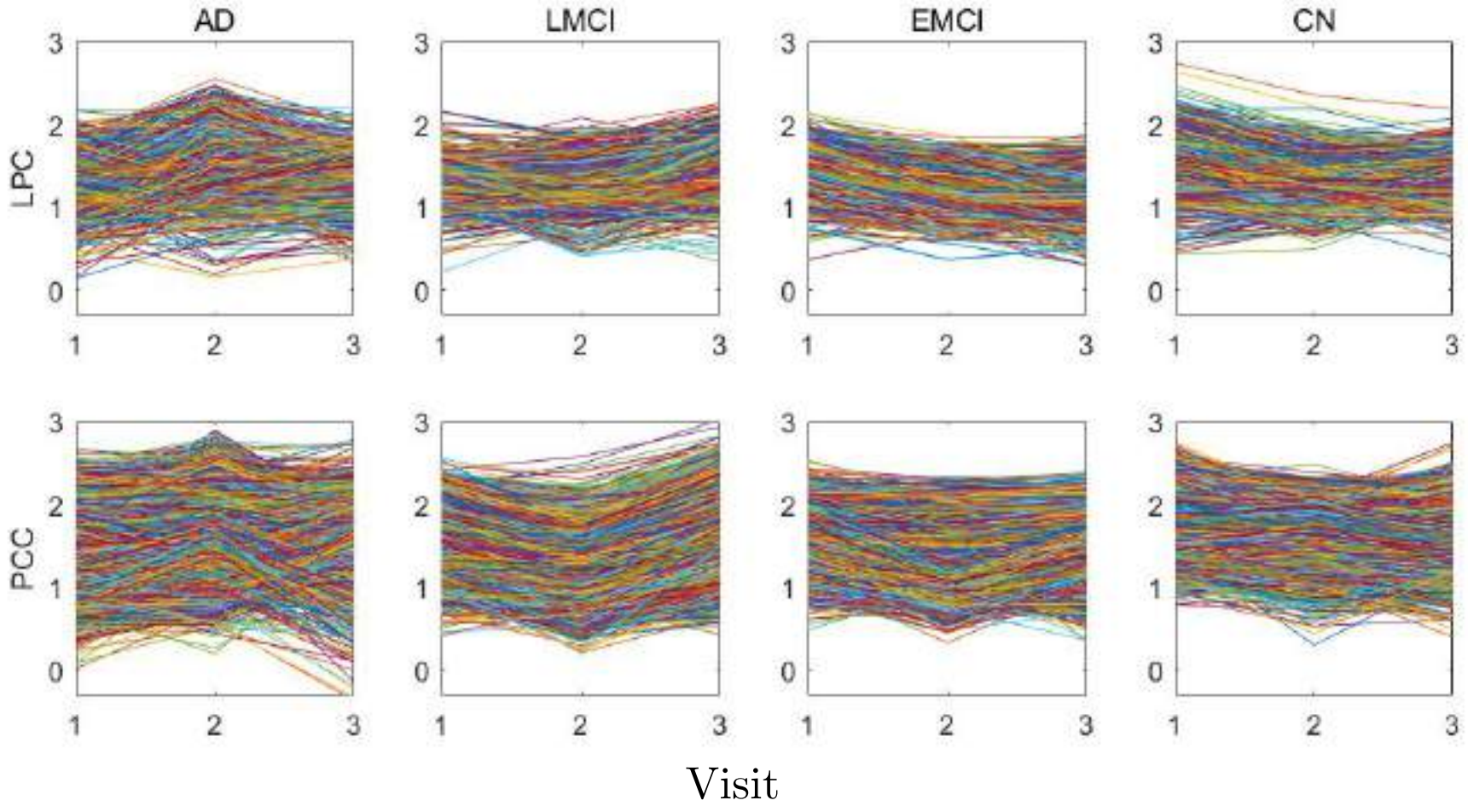

Fig 8. *L-ICA estimates of longitudinal trends for voxels in the DMN network for each disease group in ADNI2 study. Results show that AD and late MCI (LMCI) patients generally have more changes across visits and that AD group has higher within-network variations than the other disease groups at each visit.*

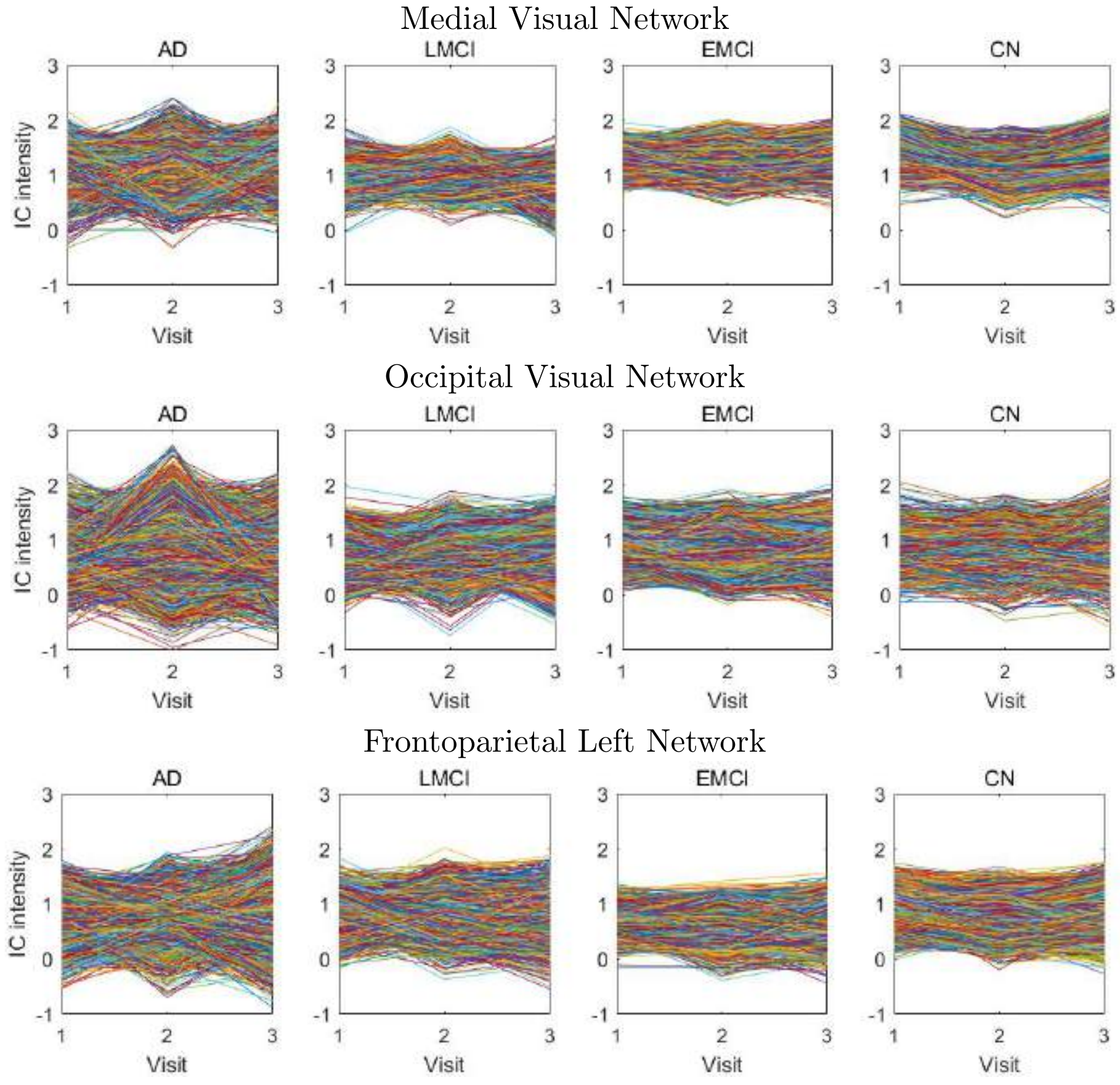

Fig 9. *L-ICA estimates of longitudinal trends for voxels in FPL and visual networks for each disease group in ADNI2 study. Results show that AD and LMCI patients generally have more changes across visits and that AD group has higher within-network variations than the other disease groups at each visit.*

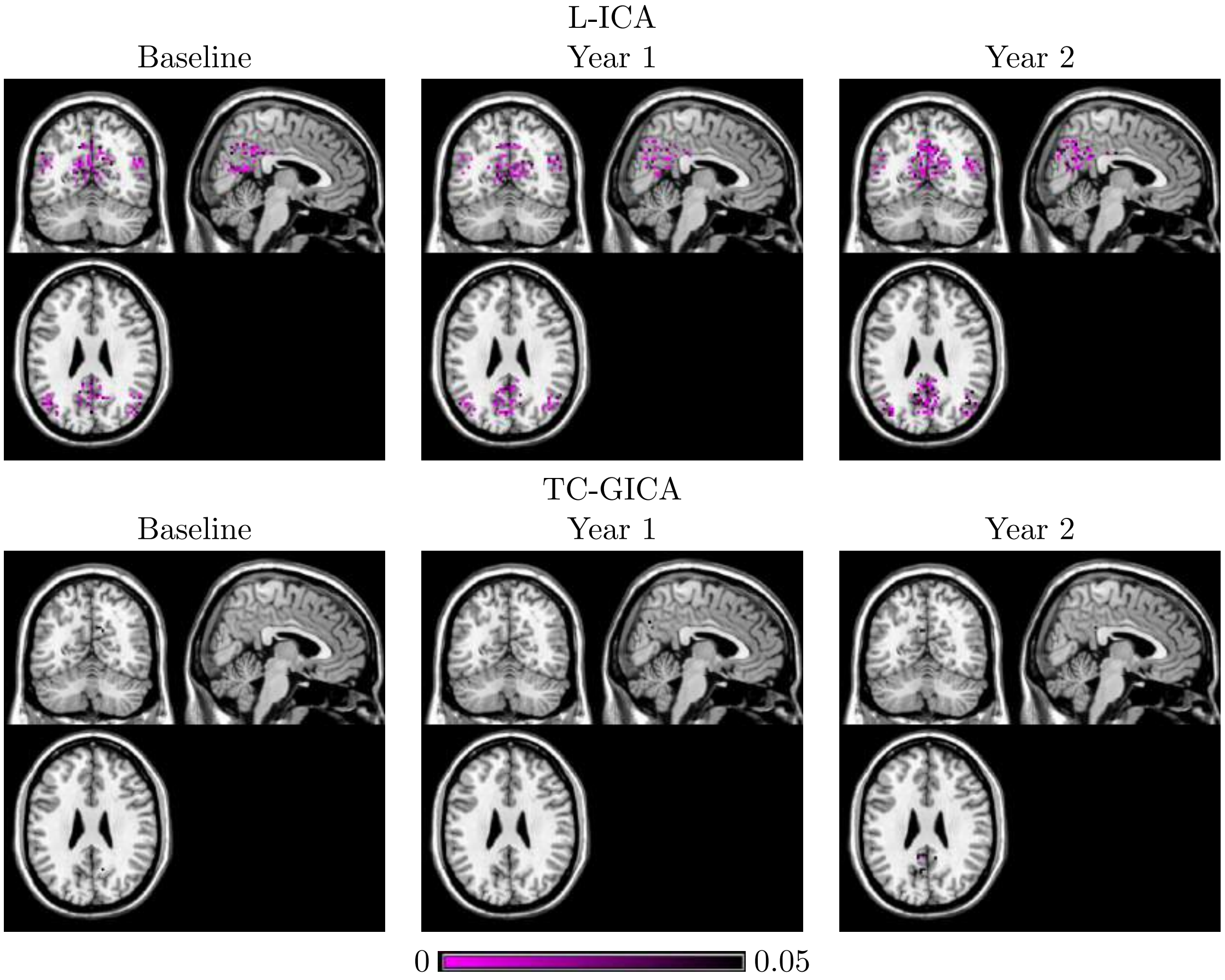

Fig 10. *p-values for testing group differences in DMN between AD and CN subjects at each visit. The first row shows the test results based on L-ICA and the second row shows the results from the TC-GICA based approach.*

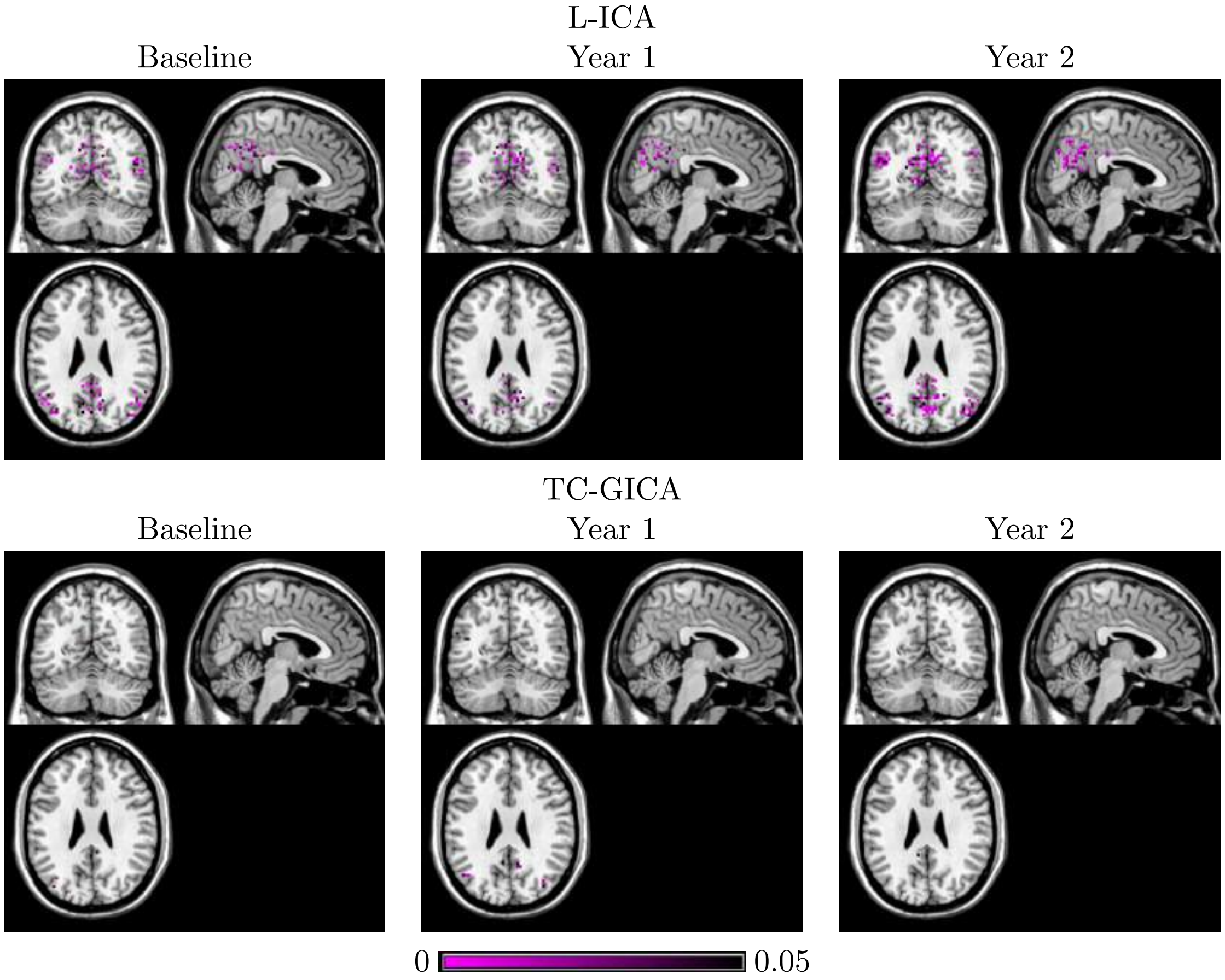

Fig 11. *p-values, thresholded at 0.05, for testing group differences in DMN between EMCI and LMCI subjects at each visit. L-ICA finds between-group differences in DMN at each visit while TC-GICA detects little group differences.*

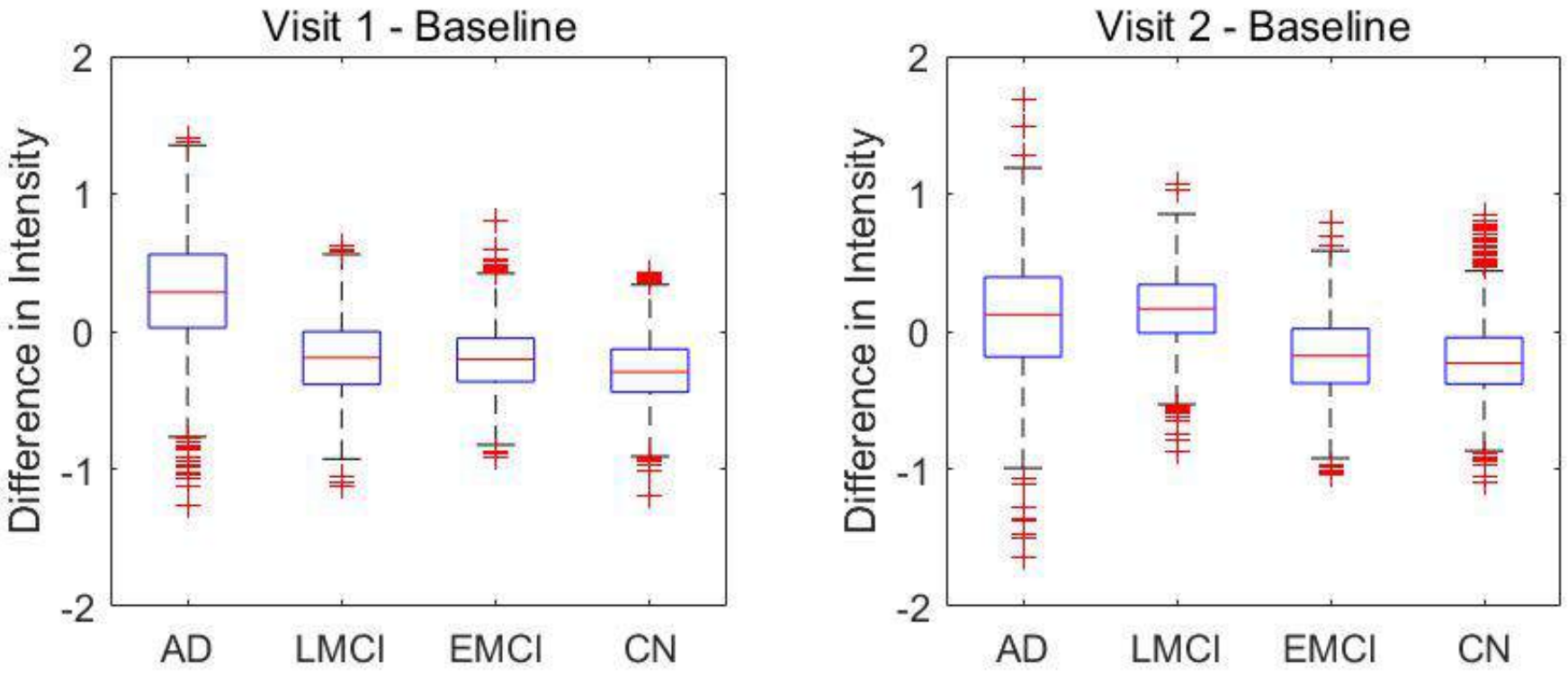

FIG 12. *Longitudinal changes from baseline and later visits in DMN within AD, LMCI, EMCI and CN groups. The first column shows the comparison between year 1 versus baseline and the second column shows the comparison between year 2 versus baseline, where the value represents the longitudinal differences in source signal intensity for DMN voxels, i.e.* $\hat{\boldsymbol{s}}_j(v) - \hat{\boldsymbol{s}}_0(v)$.

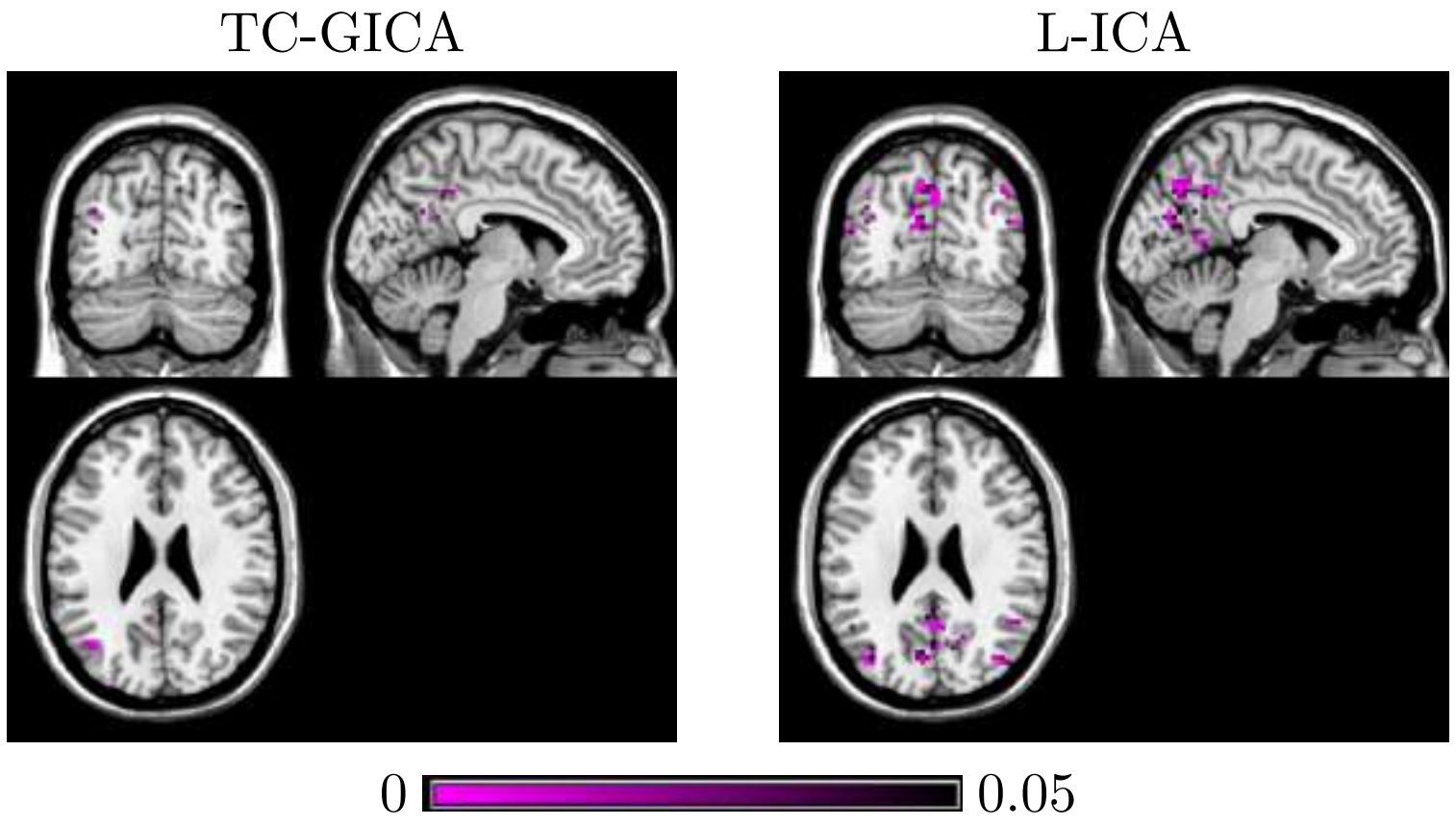

FIG 13. *p-values, thresholded at 0.05, for longitudinal changes between baseline and year 2 for the default mode network (DMN) among the AD group. L-ICA finds longitudinal changes in major regions of DMN among AD patients while TC-GICA detects little changes in DMN among these patients.*